**Assessing Dynamic Connectedness in Global Supply Chain Infrastructure Portfolios: The Impact of Risk Factors and Extreme Events**

Haibo Wang
Division of International Business and Technology Studies, A.R. Sánchez Jr. School of Business, Texas A&M International University, Laredo, TX, USA, hwang@tamiu.edu

# Abstract

This paper analyses the risk factors around investing in global supply chain infrastructure: the energy market, investor sentiment, and global shipping costs. It presents portfolio strategies associated with dynamic risks. A time-varying parameter vector autoregression (TVP-VAR) model is used to study the spillover and interconnectedness of the risk factors for global supply chain infrastructure portfolios from January 5th, 2010, to June 29th, 2023, which are associated with a set of environmental, social, and governance (ESG) indexes. The values of hedge ratio (HR) and hedging effectiveness (HE) indicate the long and short positions of investment strategies for global supply chain infrastructure portfolios. The effects of extreme events on risk spillovers and investment strategy are calculated and compared before and after the COVID-19 outbreak. The results of this study demonstrate that risk shocks influence the dynamic connectedness between global supply chain infrastructure portfolios and three risk factors and show the effects of extreme events on risk spillovers and investment outcomes. Portfolios with higher ESG scores exhibit stronger dynamic connectedness with other portfolios and factors. Net total directional connectedness indicates that West Texas Intermediate (WTI), Baltic Exchange Dry Index (BDI), and investor sentiment volatility index (VIX) consistently are net receivers of spillover shocks. A portfolio with a ticker GLFOX appears to be a time-varying net receiver and giver. The pairwise connectedness shows that WTI and VIX are mostly net receivers. Portfolios with tickers CSUAX, GII, and FGIAX are mostly net givers of spillover shocks. The COVID-19 outbreak changed the structure of dynamic connectedness on portfolios. The mean value of HR and HE indicates that the weights of long/short positions in investment strategy after the COVID-19 outbreak have undergone structural changes compared to the period before. The hedging ability of global supply chain infrastructure investment portfolios with higher ESG scores is superior. The findings of this study bring insights for investors to adjust their hedge ability to invest in global supply chain infrastructure, considering risk factors and extreme events such as the COVID-19 outbreak.

# 1. Introduction

Physical and digital infrastructures are undoubtedly the most important assets for world economic development. Investment in infrastructure is facing scrutiny from a global supply chain security perspective. Table 1 shows the rank of the top 10 countries and attribute scores regarding infrastructure in 2022. USNEWS gives the weights of attributes[1]. These attributes represent the environmental, social, and governance dimensions (ESG). Newell and Peng (2008) report the performance of investment on infrastructure in the United States for global institutional investors and provide the context of infrastructure and the rank of 17 countries in 2006.

**[Insert Table 1 here]**

Figure 1 shows the performance of four global supply chain infrastructure funds with high ESG scores in the past decade. There are downward V-shaped trends. The overall trend is upward, the big V shape in 2020 might be related to the COVID-19 outbreak. Researchers have studied the impact of COVID-19 on green financing (Pham and Nguyen 2021, Tu et al. 2021, Urom et al. 2021, Dogan et al. 2022, Mzoughi et al. 2022, Tiwari et al. 2022).

[1] https://www.usnews.com/news/best-countries/articles/methodology

Although the returns on global supply chain infrastructure investment are promising to rise in the future, some challenges might deter the sustainable development of global supply chain infrastructure as an investment asset. The funds in the study by Newell and Peng (2008) were removed from the stock exchange after 2009. Figure 1 shows four sizeable global supply chain infrastructure funds underperforming the S&P 500 from 2010 to 2023.

**[Insert Figure 1 here]**

Many risk factors should be considered when investing in global supply chain infrastructure, such as the energy market, global shipping costs, and investor sentiment. There is a rich body of literature on sustainable development, such as green bonds, but more studies on global supply chain infrastructure portfolios are needed. The impact of the ESG factors still needs to be clarified, and further analysis of their interconnection with global supply chain infrastructure investment is required. This paper first investigates the interconnections between global supply chain infrastructure portfolios and risk factors. Then, it discusses the hedging ability of global supply chain infrastructure portfolios on the dynamic risks. Therefore, this study proposes four research questions (RQs) to address the knowledge gap in the literature and investigate the interconnection between dynamic risks of global supply chain infrastructure portfolios and investment strategies.

RQ1. How does investor sentiment impact global supply chain infrastructure portfolios' returns?

RQ2. How do global shipping costs impact global supply chain infrastructure portfolios' returns?

RQ3. What is the impact of the fluctuation of the energy market on global supply chain infrastructure portfolios' returns?

RQ4. What is the impact of the COVID-19 outbreak on global supply chain infrastructure portfolios' returns?

The contributions of this study are, first and foremost, collating and analyzing data from different sources related to global supply chain infrastructure portfolios, the energy market, investor sentiment, and global shipping costs. Second, it combines investment strategy, investor behavior, and transportation costs for a comprehensive investment analysis. Thirdly, the time-varying analysis used on large-scale macroeconomic time series provides a dynamic, systematic view of the effect of global supply chain infrastructure portfolios.

The rest of the paper is structured as follows. Section 2 comprehensively reviews global supply chain infrastructure portfolios and econometric models. Section 3 proposes a TVP-VAR framework and collates data. Section 4 reports the empirical results. Section 5 discusses the findings and managerial implications. The concluding remarks are provided in section 6.

## 2. Literature Review

### 2.1 *Infrastructure portfolios as sustainable investment assets*

In recent years, infrastructure investments have gained popularity to increase returns and diversify portfolios to avert risk, driven by the growth of listed and unlisted infrastructure funds. Research has consistently shown that infrastructure investments can deliver strong risk-adjusted returns and provide significant diversification benefits when combined with other major asset classes. For example, a study by Newell and Peng (2008) found that US infrastructure investments from 2000 to 2006 outperformed other asset classes. Bamidele Oyedele (2014) discovered that UK-listed infrastructure investments offered superior risk-return trade-offs compared to property, private equity, hedge funds, and stocks. Oyedele et al. (2014) found that incorporating infrastructure investments into a mixed-asset portfolio can enhance diversification and reduce risk rather than solely boosting returns.

However, infrastructure investments also come with unique challenges and risks. Bird et al. (2014) identified regulatory risk as a key concern, while Thacker et al. (2019) highlighted the potential social and environmental impacts of poorly developed infrastructure. Hallegatte et al. (2019) estimated that delaying infrastructure investment could cost nearly $1 trillion by 2030. To address these challenges, Studart and Gallagher (2018) proposed the creation of a global guarantee fund to empower developing countries to self-finance their sustainable development initiatives.

Infrastructure investments can be valuable to a portfolio despite these challenges, particularly during economic crises. Duclos (2019) found that infrastructure investments proved more resilient than diversified portfolios during economic stress. Gupta and Sharma (2022) presented a systematic literature review on infrastructure as an asset class, highlighting its potential benefits and challenges.

Despite the growing interest in infrastructure investments as a sustainable portfolio option, several limitations and challenges must be addressed. For instance, infrastructure projects can have devastating environmental and social consequences, such as community displacement, habitat destruction, and water pollution. Private funds are not required to disclose the ESG reporting in infrastructure investments, which makes it difficult for investors to assess the project's sustainability performance and make informed decisions. Additionally, infrastructure investments often rely heavily on debt financing, which can increase the risk of default or bankruptcy, particularly in cases where the project needs to generate more cash flows. Moreover, infrastructure investments can be exposed to energy market risks, such as price volatility and supply disruptions, worsened by geopolitical conflicts and regional instability. To overcome these limitations, this study presents a framework that prioritizes ESG considerations and risk hedging capability.

2.2 *Economic models on infrastructure portfolios risk analysis*

There is a rich literature on infrastructure portfolio risk analysis using economic models. Table 2 summarizes the models and research objectives. Public capital has been extensively studied, followed by private capital and productivity. These studies provide insight into implementing these models to analyze the relationship between infrastructure investment and the socio-economic risk factors that challenge global supply chain infrastructure development.

**[Insert Table 2 here]**

The VAR model is the primary research method used to analyze the relationship between infrastructure investment and socio-economic risk factors, followed by the Cobb-Douglas production function approach and GMM. For instance, Agenor et al. (2005) noted that the poor economic performance of the Middle East and North Africa (MENA) region is mainly due to the slow pace of structural reforms and the dominant role of governments in the economy. Despite some downsizing efforts, the region's public sectors remain significant, and private sector development is affected by limited progress in building market-oriented institutions and integrating into the global economy. Maparu and Mazumder (2017) examined the causal link between transport infrastructure and economic development in India from 1990 to 2011. They found that economic growth precedes investment in transport infrastructure, supporting Wagner's law. Okolo et al. (2018) investigated the impact of capital expenditure on infrastructural development in Nigeria from 1970 to 2017 using an autoregressive distributed lag (ARDL) model. The results showed that capital expenditure, construction expenditure, and non-oil revenue can drive long-term infrastructural development but are affected by external debt, highlighting the need to boost non-oil revenue and reduce recurrent expenditure. Na (2023) found that content providers' investments in infrastructure are linked to network operators' investments and suggested that paid peering contracts are necessary for the continued growth of internet-related industries. Umar et al. (2023) examined the benefits of including listed infrastructure investments in a portfolio, using a multi-asset allocation framework to analyze short-term and long-term performance. The results showed that adding infrastructure to a traditional portfolio of stocks, bonds, and bills can benefit investors with varying levels of risk aversion, both in the short and long term.

While the VAR model provides insights into static connectedness using stationary monthly, quarterly, and annual data, dynamic conditional connectedness results using daily data can help decision-makers capture the dynamic nature of market volatility and investor behavior. This area is limited to the literature and requires further study.

# 3. Data and Methodology

## 3.1 *Data and integration tests*

There are many portfolios with global supply chain infrastructure as key investment assets, and this study chooses the portfolios with a large market cap from January 5$^{th}$, 2010, to June 29$^{th}$, 2023. The price information of the four portfolios was collected from the Center for Research in Security Prices (CRSP). VIX represents the investor sentiment indicator, and VIX's price information is obtained from CRSP. The price information of the energy market is obtained from West Texas Intermediate (WTI), provided by the Federal Reserve Bank (FRB). The Baltic Exchange Dry Index (BDI) represents Bloomberg's global shipping costs measure. The descriptions of the variables are given in Table 3. Table 4 reports the ESG risk scores of four portfolios for global supply chain infrastructure investment. The ESG risk scores of portfolios are obtained from MSCI. We separate the sample data before and after the COVID-19 outbreak to assess the effect of extreme events, which affected four portfolios, the highest on February 20, 2020 (Figure 1).

**[Insert Tables 3 and 4 here]**

3.2 Dynamic Connectedness Framework for Global Supply Chain Infrastructure Investment Strategy

This study proposes a new risk analysis and investment strategy framework, as shown in Figure 2. There are three components in this framework: 1). connectedness analyses; 2). model validation; and 3) investment strategies.

**[Insert Figure 2 here]**

Figure 2 shows five analytical modules for the time-series data: data preprocessing and integration properties, variable and time-series model validation, connectedness network, model cross-validation, investment strategies, and hedge ability. The formulations of the five analytical modules are given in the Appendix.

The first module presents the results of descriptive statistics and stationarity tests. The stationarity test is the foundation for model design in time series analysis. The descriptive statistics reveal the integration properties of the data and variables.

Using time-varying estimation, the conditional and partial correlation based on a standard VAR model in the second module can capture the correlation changes over time. The difference between conditional and partial correlation and the multivariate distribution reveals the nonlinear relationship among the variables (Lawrance 1976).

The pairwise cointegration analysis in the second module provides portfolios' cumulative return spillovers and measures the interconnection relationship between a pair of variables. Figure 3 gives the structure of the cointegration analysis among variables.

**[Insert Figure 2 here]**

There are 6 sets of hypotheses in Figure 3

H1. Strong bidirectional cointegration exists between investor sentiment and global supply chain infrastructure portfolio returns.

H2. Strong bidirectional cointegration exists between the energy market and the returns of global supply chain infrastructure portfolios.

H3. Strong bidirectional cointegration exists between global shipping costs and global supply chain infrastructure portfolios' returns.

H4. Strong bidirectional cointegration exists between investor sentiment and the energy market.

H5. Strong bidirectional cointegration exists between investor sentiment and global shipping costs.

H6. Strong bidirectional cointegration exists between the energy market and global shipping costs.

The connectedness network analyses in Module 3 include static and dynamic connectedness models. To investigate the dynamic conditional connectedness among global supply chain infrastructure, the energy market, investor sentiment, and global shipping costs in global supply chain infrastructure investment, standard VAR and TVP-VAR estimation models were performed in this study. The key difference between these two models is the type of coefficient measured by the models on shock spillover. In the case of the standard VAR model, the shock to the variable $i$ is calculated as fixed coefficients. TVP-VAR reports the shock as coefficients varying over all the periods and the variance of the estimation errors as stochastic volatility (Diebold and Yilmaz 2009). In addition, the standard VAR model estimates the time-series data based on linearity and time invariance assumptions. TVP-VAR model estimates with the time-series data containing nonlinearity.

A quantile VAR (QVAR) model is often deployed as the benchmark model of the TVP-VAR model to cross-validate the dynamic networks formed by the TVP-VAR model (Antonakakis et al. 2019). Schüler (2014) provided the theoretical background and implementation of QVAR. In addition, different rolling window intervals are used for the robustness test of the TVP-VAR model according to the literature (Antonakakis et al. 2020). Thus, there are two procedures in Module 4: The QVAR model is evaluated across quantiles, and the NET value of each variable and the TCI value of all variables in time series are represented in graphic results; the TVP-VAR model is evaluated across the different rolling window and the TCI values of different rolling window in time-series are described in graphic results.

After we conduct the risk analysis using the TVP-VAR model and draw the connectedness network to visualize the risk spillover, hedge ratios, and hedging effectiveness analysis, the last module of the framework can provide meaningful insights on how investment strategy could hedge the investment risks for global supply chain infrastructures.

In addition to analyzing hedging strategies among portfolios and external risk factors, this study investigates the impact of extreme events on connectedness and their effects on hedge ratios and hedging effectiveness. The COVID-19 outbreak is an example to illustrate the impact of such extreme events. Furthermore, this study develops dynamic networks of risk spillover and hedging strategies for four portfolios and external factors, providing a systematic framework for optimizing global supply chain infrastructure investments.

# 4. Empirical Results

## 4.1 *Data preprocessing*

Table 5 reports the descriptive statistics for the first-order difference of the four global supply chain infrastructure investment portfolios, VIX, WTI, and BDI. Table 5 shows that the medians of all variables are positive and close to zero, with CSUAX having the highest value. GLFOX has the most significant standard deviation value among the four portfolios, indicating GLFOX exhibits more volatility than other portfolios. All variables except BDI are negatively skewed, indicating that the probability density function has a longer left tail. BDI is positively skewed, indicating a longer right tail. BDI has a skewed value close to zero with the shape of a standard normal distribution. On the other hand, the high kurtosis values indicate that these variables have fat tails in the distribution. Jarque-Bera (JB) normality test rejects the null hypothesis of normality for all variables at a 0.5% significance level (Jarque and Bera 1980). The results of the Ljung-Box Q test show that all variables have a serial correlation in squared series, suggesting that each variable has time-varying variance (Ljung and Box 1978). Thus, the TVP-VAR model is appropriate for measuring the connectedness network among the variables.

**[Insert Table 5 here]**

In addition to the descriptive statistics, a test of the integration properties of the variables is performed using an Augmented Dickey-Fuller (ADF) unit-root test before further analysis. The ADF test results in Table 6 revealed that portfolios, investor sentiment, WTI, and BDI are stationary at the first-order difference in all samples; these variables are integrated into order one. The first-order difference form of the variables should be used in further analysis.

**[Insert Table 6 here]**

4.2 *VAR Conditional and partial correlation*

Using the VAR model, tables 7 and 8 show the conditional and partial correlation between four global supply chain infrastructure portfolios, investor sentiment, energy market, and global shipping costs. The conditional correlation in Table 7 indicates that investor sentiment and global shipping costs negatively affect global supply chain infrastructure portfolios. The relationships between global supply chain infrastructure portfolios and the energy market are positive. The partial correlation in Table 8 indicates that investor sentiment, energy market, and global shipping costs have mixed and weak relationships with global supply chain infrastructure portfolios. The differences between conditional and partial correlation indicate that the relationship between these variables might be nonlinear.

**[Insert Tables 7 and 8 here]**

A CHOW test is performed to evaluate the nonlinear relationship further, and the parameter stability results are reported in Table 9. The null hypothesis of the CHOW test is rejected with a p-value of 0.008. Thus, the breakpoint test shows the estimated parameters' instabilities in the linear VAR model. After the CHOW test detects the structural instabilities, the standard linear VAR model might not capture the nonlinear relationship among variables. The CHOW test supports the TVP-VAR model as a better method for the nonlinearity of stochastic volatility.

**[Insert Table 9 here]**

4.3 *Integration properties*

If there is a nonlinear relationship among the variables, then a standard VAR can be used to assess the interconnection relationships when variables have pairwise cointegration relationships. Thus, this study performed the augmented Engle-Granger two-step cointegration test, and the pairwise cointegration test results are presented in Table 10. All variables have a solid bidirectional cointegration relationship. The Jarque-Bera (JB) goodness-of-fit test in Table 5 shows the distribution of residuals of the cointegration relationship for the variables. Thus, all hypotheses (H1-H7 in section 3.2) are supported by evidence. *P*-values <0.05 indicate the validity of the cointegration analysis for variables. Pairwise cointegration between two variables implies the existence of system-wide cointegration.

**[Insert Table 10 here]**

4.4 *Connectedness network*

To assess the impact of the energy market, investor sentiment, and global shipping costs on global supply chain infrastructure portfolios (RQ1 to RQ3), a set of tests for dynamic connectedness are used to estimate the spillover contributions on global supply chain infrastructure portfolios.

Since the variables are pairwise cointegrated, a VAR connectedness with time was performed for portfolio connectedness analysis. The results are reported in Table 11. Three global supply chain infrastructure Portfolios, namely GII, FGIAX, and CSUAX, are the primary givers of spillover shocks in the financial network. FGIAX has the highest contribution, and CSUAX has the lowest contribution.

**[Insert Table 11 here]**

WTI, VIX, GLFOX, and BDI are the receivers of spillover shocks in the financial network, and WTI receives the highest influence from others. In Table 11, the diagonal elements represent their variable spillovers of shocks, and the rest of the elements capture the interaction across variables. The average TCI in the static connected network is 47.52%, indicating that the shock of all other variables can explain 47.52% of the forecast error variance of one variable. The "NET" row means that WTI appears to be the largest net giver with a NET value of -13.83%. Thus, WTI is likely to be affected by other variables. In contrast, FGIAX is the largest net giver in the network, with an average net connectedness value of 11.42%. FGIAX, GII, and CSUAX are likely to affect other variables. The graphic results of static connectedness are offered in Figure 4, where the nodes in blue are net givers, and the nodes in yellow are net receivers of shocks in the financial network.

**[Insert Figure 4 here]**

The arrow in the graph indicates greater directional connectedness in terms of net connectedness. The size of the nodes in the graph shows the weighted average net total directional connectedness in the financial network. The bold lines in the graph indicate a more significant spillover than fine lines between variables. The bold line between GII and WTI indicates a more substantial spillover from GII to WTI; there are bold lines on other pairs of nodes, such as from GII to GLFOX, from GII to VIX, from CSUAX to WTI, from FGIAX to CSUAX, from FGIAX to VIX, from FGIAX to WTI, from FGIAX to GLFOX. Thus, there are higher contributions on the portfolios with high ESG scores. However, there is no bold line between BDI and other variables. One explanation is that global shipping costs might not have a linear relationship with different variables, and VAR connectedness analysis cannot capture nonlinear relationships. As mentioned earlier, VAR connectedness analysis produces reliable results when the variables have linear relationships, but conditional and partial correlation analysis in section 4.2 found nonlinear relationships among variables.

Although the results of the static connectedness measures from VAR provide an overview of the underlying interrelations in the financial network when variables have linear relationships, a framework using the TVP-VAR model can offer more insights into the time-varying connectedness among the variables with nonlinear relationships.

Regarding the dynamic conditional connectedness, this study reports the results using the TCI to measure each variable's average impact on all other variables. The dynamic conditional connectedness analysis results in Table 12 provide a good picture of the volatility connectedness between four global supply chain infrastructure portfolios, energy market, investor sentiment, and global shipping costs. All global supply chain infrastructure portfolios are the primary givers of shocks in the financial network, while FGIAX has the highest contribution, and GLFOX has the lowest contribution. VIX, WTI, and BDI are the net receivers of spillover shocks in the financial network, and VIX receives the highest influence from others. Thus, VIX strongly affects global supply chain infrastructure portfolio returns, and shocks of return spillovers and volatility in returns affect the energy market and global shipping costs. This might explain the 2016 downward trend of global supply chain infrastructure portfolios associated with the oil price decline and investor sentiment.

**[Insert Table 12 here]**

The value of TCI in Table 12 shows that the average dynamic conditional connectedness value in the financial network is 57.32%, which indicates that the shock spillovers of all other variables can explain the forecast error variance of one variable within this financial network. The value of TCI for dynamic conditional connectedness is higher than that of TCI's for static connectedness from the VAR model. VIX appears to be the largest net receiver based on the NET value, with an average net connectedness value of −18.26%, which means VIX is likely to receive shocks from the other variables. Meanwhile, the negative sign of the net spillover can also be found in WTI and BDI, with values of -15.63% and -9.18%. By contrast, FGIAX is the largest net giver in the financial network, with a net connectedness value of 16.95%. The positive sign of the net spillover occurs in CSUAX, GII, and GLFOX and indicates that these variables are net givers of shocks. Compared to the static connectedness from the VAR model in Table 11, GLFOX in the TVP-VAR model is the net giver of a shock instead of the receiver in the VAR model. The graphic representations of the dynamic connectedness among four portfolios and three factors are given in Figures 5, 6, 7, and 8.

**[Insert Figures 5, 6, 7, and 8 here]**

The bold line in Figure 8 between GII and VIX indicates more significant spillovers from GII to VIX. There are bold lines such as from GII to WTI, from GLFOX to VIX, from FGIAX to VIX, from FGIAX to WTI, from FGIAX to GLFOX, from CSUAX to VIX, and from CSUAX to WTI. There are more bold lines in Figure 8 than in Figure 4. We find the VIX having strong relations with GLFOX in the TVP-VAR model, which is nonexistent in the VAR model. There are bold lines from four portfolios to VIX and from three portfolios to WTI. Figure 8 shows that VIX, WTI, and BDI are net receivers of shocks. Four portfolios are net givers of shocks. This dynamic connectedness network's members and ranks from the TVP-VAR model help us better understand the hedge's ability to manage dynamic risks.

4.5 *Cross-validation*

This study validates the findings of the TVP-VAR model using the QVAR model across a sequence of quantiles with a 10% interval as follows: q=0.05, 0.15, 0.25, …,0.95. The results of QVAR dynamic connectedness are mapped in heatmap graphs in Figure 9. The dynamic connectedness from 2010 to 2023 for the TCI value of all variables and the NET value of individual variables is represented by different shades. The legend of the heatmaps shows strong dynamic connectedness with warmer shades and weak connectedness with colder shades. The title of each heatmap, except the first one, indicates the NET values of the corresponding variable. The first heatmap for the TCI value of all variables shows that the upper quantiles (above the 75$^{th}$ quantile) correspond to positive market conditions. In contrast, lower quantiles (below the 25$^{th}$ quantile) represent negative market conditions from 2010 to 2023, having strong dynamic connectedness, and medium quantiles have weak dynamics in general, except during the years 2012, 2016, and 2020 with deep warmer shades. The results indicate symmetric spillover effects without extreme events like the COVID-19 outbreak. The heatmap of TCI shows strong dynamic connectedness during 2020 across all quantiles. Thus, the COVID-19 outbreak has changed the spillover effects on all variables, and it motivates us to explore the hedging ability in the following subsection.

**[Insert Figure 9 here]**

Other heatmaps in Figure 9 show the net directional connectedness of individual variables and the sensitivity of NET values across quantiles from 2010 to 2023. The warmer shades represent the net giver, and the colder shades represent the net receiver. FGIAX and CSUAX are primarily net givers of shock in the TVP-VAR model (see Figure 8), and both have warmer shades between the $75^{th}$ quantile and $25^{th}$ quantile, indicating normal market conditions throughout the entire period in the QVAR model, but much warmer shades during the years 2016 and 2020. In the TVP-VAR model, GII and GLFOX are net givers of shock but have less influence on spillovers than FGIAX and CSUAX. The rank of net givers of the TVP-VAR model in Table 12 for FGIAX, CSUAX, GII, and GLFOX can be observed from the heatmaps of the QVAR model across quantiles. FGIAX has a larger area of warmer shades than CSUAX after 2016. The warmer shades in CSUAX are more significant than those in GII, which had colder shades before 2016. GLFOX has mixed warmer and colder shades across the quantile in the QVAR model during the entire period, which indicates that GLFOX could switch from one net shock behavior to another during the whole period. GLFOX has the lowest net giver value in the TVP-VAR model.

The rank of net receivers of the TVP-VAR model in Table 12 for VIX, WTI, and BDI can be observed from the QVAR heatmaps in Figure 9. VIX has a larger area of colder shades than other variables, followed by WTI. VIX has deep, colder shades between the $75^{th}$ and 25th quantiles after the COVID-19 outbreak 2020. WTI has light coder shades between the 75th and 25th quantiles. BDI has colder shades, and GLFOX has an interesting pattern of switching from colder shades to deep, warmer shades after the COVID-19 outbreak in 2020, which suggests a strong effect of COVID-19 on the net shock behavior; GLFOX switches from a net receiver to a strong net giver in a brief period. We observed that light, warmer, and colder shades existed in heatmaps for all variables. This heterogeneous net dynamic connectedness could indicate the variables' sensitivity to extreme events, and the variable can switch from net receiver to net giver based on the external circumstance. This provides evidence for addressing the connection between COVID-19 and spillovers in RQ4. We find that the results of QVAR heatmaps validate the findings of the TVP-VAR model on the member and rank of net givers and net receivers in Table 12.

4.6 *Robustness test*

After validating the TVP-VAR model on the member and rank of dynamic connectedness of variables, this study performs a robustness test of the TVP-VAR model to further examine the sensitivity of the findings according to the choice of the lengths of the windows. This study examines the net directional accuracy of each variable using TVP-VAR across rolling windows of 60, 120, 180, and 360 days to evaluate the robustness of the dynamic conditional connectedness analysis among these variables. These results are reported in Table 13. One variable's net directional spillover contribution on the other variable is stable across any given rolling window. Figure 10 shows the NPDC measure plots across rolling windows of 60-day, 120-day, 180-day, and 360-day. The size of the node and the number of bold lines vary across rolling windows. The primary net givers of shock, FGIAX, CSUAX, GII, and GLFOX, have greater spillover contributions across rolling windows. VIX is the largest receiver across all rolling windows, followed by WTI. The standard deviation of the spillover contribution from givers is greater than from receivers in Figure 11, which confirms a robust dynamic interconnection between spillover and volatility connectedness within the system. The blue line has higher values than the purple line. The shapes of both lines are similar over time.

**[Insert Figures 10, 11, 12, 13, and 14 here]**

**[Insert Table 13 here]**

The intertemporal changes shown in Figure 15 show that the TCI values vary during the whole period. Their values sharply increased in March 2020 after a decrease from December 2020 to late February 2020. This suggests a marked connectedness between four portfolios: energy market, investor sentiment, and global shipping costs linked to the COVID-19 outbreak. The results of TVP-VAR across rolling windows of 60-, 120-, 180-, and 360-day are similar to those in Table 12 regarding members and rank of the connectedness network. This indicates that our results for the TVP-VAR model are robust.

**[Insert Figure 15 here]**

### 4.7 *The effect of extreme events on the role of shock spillover*

The findings from TVP-VAR and QVAR models indicate that the energy market, investor sentiment, global shipping costs, and extreme events can affect the dynamic risks of global supply chain infrastructure portfolios. There is a heterogeneous net dynamic spillover pattern across all portfolios, and the role of shock receiver and giver can be switched due to extreme events.

To provide a comprehensive analysis for the proposed RQ4, the TVP-VAR model is applied to the samples before and after the COVID-19 outbreak, and HR and HE are calculated for a complete set of observations on the samples before and after the COVID-19 outbreak for comparison. Figure 16 shows that GLFOX was a shock receiver before COVID-19 and became a shock giver after the outbreak. Global shopping costs and the energy market had a greater spillover effect on the global supply chain infrastructure portfolios before the COVID-19 outbreak compared to their impact after it. Figure 17 illustrates the surge of TCI value, indicating strong dynamic connectedness among the variables.

**[Insert Figures 16 and 17 here]**

The heatmap of the QVAR model in Figure 18 shows that GLFOX was a net receiver of shock from 2018 to 2020 under normal market conditions, but became a net giver of shock after the COVID-19 outbreak, as illustrated in Figure 19. The TCI heatmaps in the QVAR model and the TCI plots of the TVP-VAR model show the structural change of dynamic connectedness linked to the COVID-19 outbreak.

**[Insert Figures 18 and 19 here]**

4.8 *Hedgeability of global supply chain infrastructure portfolios vs risk factors*

Table 14 reports the values of HR and HE for the *global supply chain infrastructure* portfolios and the risk factors on the complete set of observations. The value of the HR between a long position in the global supply chain infrastructure portfolio and a short position in global shipping costs (BDI) is the lowest, indicating that the cheapest hedge for a $1 long position in FGIAX is obtained with BDI (6 cents), followed by a long position in GLFOX with BDI (7 cents), a long position in CSUAX with BDI (7 cents), and a long position in GLFOX with BDI (10 cents). The value of the HR between a long position in the global supply chain infrastructure portfolio and a short position in the energy markets (WTI) is higher than that for global shipping costs (BDI), with 43 cents for GII, 40 cents for GLFOX, 47 cents for FGIAX, and 41 cents for CSUAX. The negative HR ratio between a long position in the global supply chain infrastructure portfolio and a short position in

investor sentiment, as measured by the VIX, indicates that the VIX moves in the opposite direction to the global supply chain infrastructure portfolio, suggesting an aggressive hedging strategy by investors. In practice, this hedge strategy for a long position in the global supply chain infrastructure portfolio indicates that when the prices of the global supply chain infrastructure portfolio decrease, there are opportunities to buy put options or sell call options to reduce risk and increase the value of the holdings.

The hedge effectiveness for global supply chain infrastructure portfolios in the long position and risk factors in the short position is low, indicating that investors will incur high costs to hedge risks. The hedge effectiveness for global supply chain infrastructure portfolios in the long position and VIX in the short position is higher than for BDI and WTI, suggesting that investors can hedge more risks with VIX at a lower cost. Global supply chain infrastructure portfolios with higher ESG scores, such as GII and CSUAX in long positions and VIX in short positions, exhibit higher hedge effectiveness, indicating a more effective hedging ability. However, the negative HE values for BDI in a short position indicate that hedging with BDI is ineffective despite the low cost of hedging.

**[Insert Table 14 here]**

Table 15 reports the value of HR and HE for portfolios before and after the COVID-19 outbreak. There have been profound changes in the hedge ratio and hedging effectiveness since the COVID-19 outbreak. The hedge ratios are significantly higher, indicating a higher cost to hedge risks in global supply chain infrastructure portfolios when taking a long position in those stocks.

When considering global supply chain infrastructure portfolios with a short position and risk factors with a long position, the hedge effectiveness decreases, indicating the hedging ability worsens after COVID-19. This suggests that a short position in global supply chain infrastructure portfolios and a long position in three factors will not effectively reduce the risk.

Combining the results of Table 14, the findings of HR and HE in Table 15 provide evidence for structural changes of dynamic risks due to extreme events. These findings suggest that investors adjust the hedging ability of the portfolio according to the circumstances.

**[Insert Table 15 here]**

# 5. Discussion

This study presents a TVP-VAR framework to estimate the interconnectedness of four global supply chain infrastructure portfolios: energy market, investor sentiment, and global shipping costs. QVAR validates the results of interconnectedness across quantiles. TVP-VAR and QVAR found the same member and rank of the dynamic connectedness network on the complete set of observations. However, the effects of external factors on four global supply chain infrastructure portfolios are time-varying. The hedge ability of portfolios and external factors pairs is sensitive to extreme events. The impacts of external factors on portfolios are discussed in the following subsections.

## *5.1 Impact of the energy market*

The findings from this study show that the energy market is the primary net receiver of the shocks from global supply chain infrastructure investment. Russia's invasion of Ukraine has tightened the supply chains in the energy market and pushed up the prices in global energy markets. There is an immediate need to diversify the supply chain and a long-term need to develop the global supply chain infrastructure to improve the resiliency of the global supply chain. However, the rising inflation and higher labor and materials costs make investors and companies hesitate to increase their spending due to the foreseeable risks. The energy market has positive relationships with the returns of four global supply chain infrastructure portfolios in terms of conditional correlation coefficients, and the relationships are stronger than global shipping costs with the returns of the portfolios. Thus, the returns of global supply chain infrastructure portfolios affect the energy market and interact with each other. The hedging effectiveness of the global supply chain infrastructure investment portfolio in a long position and the energy markets in a short position is significantly higher than that of global shipping costs and VIX in a short position. Additionally, the hedging effectiveness of global supply chain infrastructure investment portfolios with higher ESG scores in a long position and the energy markets in a short position is much higher than that of portfolios with lower ESG scores. This result aligns with the empirical evidence in the literature, which suggests that portfolios with higher ESG scores can lead to better risk-adjusted returns (Verheyden et al. 2016, Daugaard 2020, Shanaev and Ghimire 2022). The hedge effectiveness for global supply chain infrastructure portfolios in a long position and the energy market in a short position increased after the COVID-19 outbreak.

5.2 *Impact of investor sentiment*

The investor sentiment index VIX measures the fear and greediness of investors in the stock market, as well as their economic behavior. It negatively affects the returns of four global supply chain infrastructure portfolios regarding conditional correlation coefficients. The relationships are stronger than the energy market and global shipping costs with portfolio returns. TVP-VAR conditional connectedness shows that investor sentiment is the primary receiver of shocks in the financial network. Thus, the returns of portfolios affect investors' behavior, and they interact with each other. The results of TVP-VAR and QVAR show that the dynamic connectedness between investor sentiment and global supply chain infrastructure portfolios before the COVID-19 outbreak is stronger than the dynamic connectedness after.

In addition, the hedging effectiveness of global supply chain infrastructure investment portfolios with higher ESG scores in a long position and VIX in a short position is higher than that of portfolios with lower ESG scores. Thus, the hedging ability of global supply chain infrastructure investment portfolios with higher ESG scores is superior.

5.3 *Impact of global shipping costs*

Hendrix (2022) and Signé and Johnson (2021) highlighted the importance of global shipping costs on global supply chain infrastructure development. Investment in global supply chain infrastructure exhibits weak negative relationships with the returns of global shipping costs, as indicated by conditional correlation coefficients. These relationships are weaker than those observed with the energy market and investor sentiment.

The hedging effectiveness of global shipping costs in a long position and global supply chain infrastructure investment in a short position is significantly lower than that of the energy market in a short position. This indicates a less effective hedging ability for global shipping costs against global supply chain infrastructure investment than the energy market and investor sentiment.

5.4 *Managerial implications*

The findings of this study provide meaningful insights for companies, investors, and regulators on systematic risk and investment strategies. This paper is the first to present a comprehensive discussion of the interconnections between global supply chain infrastructure investment and risk factors and the hedging ability of portfolios on dynamic risks. The structural changes before and after COVID-19 suggest that investors and regulators should be aware of the contagion influences of extreme events and uncertain risks from other markets.

The spillovers from other markets can affect price fluctuations and short-term volatility in global supply chain infrastructure portfolios. Collecting quantitative data from different markets on the same time scale or period is challenging. Any change in international regulatory policies might bring more complexity to the systematic shock, and some of the shocks might not be captured by the existing models or methods directly. New models or measurements might consider these risks for data collection and model design in our future paper.

Strategic investment in transportation, public buildings, and critical infrastructure creates jobs and drives economic growth. Rather than relying on tax cuts and corporate incentives, states should prioritize investments in infrastructure that lay the groundwork for a strong economy. The condition of a state's infrastructure has a direct impact on its economic performance. A well-maintained infrastructure is essential for commerce to thrive, enabling manufacturers in the global supply chain ecosystem to access raw materials and deliver products to consumers efficiently. By investing in infrastructure improvements, states can unlock economic benefits, including increased private-sector investment and productivity growth.

# 6. Conclusions

The results of the TVP-VAR approach indicate that a subset of global supply chain infrastructure investment portfolios (with the tickers GII, FGIAX, and CSUAX) are the primary givers of shocks, which in turn affect the returns of other portfolios in the financial network. One portfolio (with the ticker GLFOX) acts as both a giver and receiver of shocks in the network. The three risk factors (the energy market, investor sentiment, and global shipping costs) are net receivers of return spillovers. The Total Connectedness Index (TCI) values from the dynamic conditional connectedness analysis using the TVP-VAR model provide greater explanatory power than the static connectedness analysis using the standard VAR model. The heatmaps generated by the QVAR model validate the findings of the TVP-VAR model regarding the membership and ranking of the connectedness network. Portfolios with higher ESG scores exhibit stronger dynamic connectedness with other portfolios and factors. Additionally, the QVAR heatmaps show warmer shades for portfolios with higher ESG scores.

The dynamic conditional connectedness approach reveals that the members and ranks of givers and receivers are consistent when considering them as individual factors on portfolio returns in the TVP-VAR and QVAR models. Although the values of receivers and givers differ slightly in each analysis, they collectively provide a good sensitivity analysis and robustness test assessment.

The hedge ratios and hedging effectiveness results indicate that portfolios with higher ESG scores have a more effective hedging ability. Extreme events, such as the COVID-19 outbreak, change the structure of dynamic connectedness and hedging effectiveness.

The results of our analytics framework indicate a significant level of interconnectedness between the returns of the four portfolios and three risk factors. These results confirm the presence of market risks within the financial networks, both in terms of returns and volatility, with conditional interconnectedness. The findings of this study provide meaningful insights for investors and suggest that investors should adjust their hedging strategies when investing in global supply chain infrastructure portfolios, taking into account risk factors and extreme events.

# Appendix

## *The formulation of the VAR model*

The standard VAR model is given as:

$$x_t = d_t + \sum_{j=1}^{L} \emptyset_j x_{t-j} + w_t,\ w_t \sim Normal(0, \Sigma_w) \qquad (1)$$

, where $x_t$ is the vector of variables at time *t* with a linear function including its own lags $L$, conformable matrices $\emptyset_j$representing lag dynamics, a vector of the deterministic components $d_t$, and a vector of estimation errors $w_t$. Equation (1) is valid if there is no structural change in the parameter estimations between time intervals. If there are structural breaks in the parameter estimations, then the theoretical assumption of the standard VAR is violated. This study used a CHOW test to evaluate the stability of the standard VAR model.

## *The formulation of pairwise cointegration analysis*

This study performs the pairwise cointegration analysis using the augmented Engle-Granger two-step cointegration test. (MacKinnon 1994). The Jarque-Bera (JB) goodness-of-fit test is used to validate the output of pairwise cointegration and to evaluate the distribution of residuals of the cointegration relationship between the variables in Figure 3.

$$\begin{bmatrix} inf_t \\ inv_t \\ wti_t \\ bdi_t \end{bmatrix} = d_t + \emptyset_1 \begin{bmatrix} inf_{t-1} \\ inv_{t-1} \\ wti_{t-1} \\ bdi_{t-1} \end{bmatrix} + \emptyset_2 \begin{bmatrix} inf_{t-2} \\ inv_{t-2} \\ wti_{t-2} \\ bdi_{t-2} \end{bmatrix} + \cdots + \emptyset_p \begin{bmatrix} inf_{t-p} \\ inv_{t-p} \\ wti_{t-p} \\ bdi_{t-p} \end{bmatrix} + w_t \qquad (2)$$

*where:*

$inf_t$ Global supply chain infrastructure Portfolios (GII, GLFOX, FGIAX, CSUAX)

$inv_t$ Investor Sentiment Fear Index VIX

$wti_t$ The Energy Market (WTI)

$bdi_t$ Global Shipping Costs (BDI)

*The formulation of the TVP-VAR model*

The basic TVP-VAR model is then:

$$x_t = \sum_{j=1}^{p} \emptyset_t x_{t-j} + w_t, w_t \sim Normal(0, S_t) \quad (3)$$

$$\emptyset_t = \emptyset_{t-1} + v_t, \; v_t \sim Normal(0, R_t) \quad (4)$$

, where $x_t$ is a vector of variables of interest, $p$ is the best lag length based on a Bayesian information criterion (BIC). Both $w_t$ and $v_t$ are vectors of the estimated error, while $S_t$ and $R_t$ are time-varying variance-covariance matrices and $\emptyset_t$ is a matrix for the estimated coefficients of each variable at time *t*.

The shock spillover from variable *i* to variable *j* can be measured by a variance decomposition matrix $D_{ij}(h)$ with element $d_{ij}(h)$ as the $h$-step ahead predicted error variance decompositions and $l_{ij}(h)$ is the standardized value of $d_{ij}(h)$. "Receiver" of the spillover effect is:

$$S_{i\rightarrow}(h) = 100 \frac{\sum_{j=1, i\neq j}^{N} l_{ij}(h)}{\sum_{j=1}^{N} l_{ij}(h)}. \quad (5)$$

"Givers" of the spillover effect is given as:

$$S_{\rightarrow i}(h) = 100 \frac{\sum_{j=1, i\neq j}^{N} l_{ji}(h)}{\sum_{j=1}^{N} l_{ij}(h)}. \quad (6)$$

The total connectedness of return spillover effects is given as follows:

$$TCI(h) = 100 \frac{\sum_{i,j=1, i\neq j}^{N} l_{ij}(h)}{\sum_{i,j=1}^{N} d_{ij}(h)} \quad (7)$$

and the net pairwise directional connectedness (NPDC) is given as:

$$NPDC_{i\leftarrow j} = l_{ij} - l_{ji} \quad (8)$$

The TVP-VAR measures the dynamic conditional connectedness among variables by evaluating the average impact of a shock in each variable and its volatility on all the other variables. The TVP-VAR reports a set of indexes (*TCI*) measuring the interconnectedness among the energy market, investor sentiment, global shipping costs, and investment portfolio in global supply chain infrastructure; $Receiver_i$ connectedness index represents the directional spillover received by the variable $i$ from all other variables; $Giver_i$ connectedness index represents the directional spillover sent by the variable $i$ to all other variables; The differences between the total directional $Giver_i$ and $Receiver_i$ is given as $NET_i$, which is the net influence of the variable $i$; "Net Pairwise Directional Connectedness (NPDC)" measures the influence/spillover variable $i$ has on variable $j$. If NPDC is positive, then $i$ dominates $j$, otherwise, $j$ dominates $i$. In general, the value of net spillover could suggest to investors the position in risk hedging strategies. The negative value of $NET_i$ indicates the long position of the portfolio, and the positive value indicates the short position.

For an extensive network, $NPDC_{i,j}$ are computationally efficient and more accurate than *TCI* in reducing bias results. The value and sign of the net total directional connectedness identify whether a specific variable is a net transmitter (giver) or a net receiver of uncertainty shocks over time and its rank in the network. The value and sign of $NPDC_{i,j}$ indicate the dynamic conditional connectedness between two specific variables. The net pairwise directional connectedness ($NPDC_{i,j}$) measure is also graphically reported.

*The formulation of hedge ratios and effectiveness*

Hedge ratios for minimum variance (MV) of one portfolio are implemented according to (Kroner and Sultan 1993):

$$H_{mv} = \frac{Cov(\Delta c_t, \Delta f_t)}{Var(\Delta f_t)} \tag{9}$$

where $C_t$ and $F_t$ are current and future prices at time t. $\Delta c_t$ and $\Delta f_t$ are first-order differences for $C_t$ and $F_t$: $\Delta c_t = ln(C_t) - ln(C_{t-1})$ and $\Delta f_t = ln(F_t) - ln(F_{t-1})$

For pairwise hedge ratio of two portfolios, the hedge ratio is given:

$$H^1{}_{p1,p2} = \frac{h_{p1,p2}}{h_{p1}} \tag{10}$$

where $H^1{}_{p1,p2}$ is the hedge portfolio with a one-dollar long position in the portfolio $p_1$ and a short position in the portfolio $p_2$, $h_{p1,p2}$ is the conditional covariance between portfolios $p_1$ and $p_2$. $h_{p1}$ is the conditional variance of portfolios $p_1$. The hedge ratios between infrastructure portfolios and external risk factors, namely, the energy market, investor sentiment, and global shipping costs, are reported in this module.

In addition to the hedge ratio, the study also reported hedging effectiveness. Hedging effectiveness is calculated according to (Ederington 1979):

$$HE_{p1,p2} = \frac{h_u - h_{p1,p2}}{h_u} \tag{11}$$

where $h_u$ is the conditional variance of the portfolio $p_1$ without hedging strategies. The details of implementation for hedging effectiveness are described by (Antonakakis et al. 2020).

Table 1. Rank of countries and attribute scores on infrastructure in 2022

| **Rank** | 1 | 2 | 3 | 4 | 5 | 6 | 7 | 8 | 9 | 10 |
|---|---|---|---|---|---|---|---|---|---|---|
| **Attribute Score** | Germany | Japan | USA | UK | France | South Korea | Canada | Denmark | Sweden | Switzerland |
| **Adventure** | 26.8 | 45.1 | 42.4 | 37.4 | 68.7 | 21.9 | 56.8 | 44.6 | 55.1 | 61.5 |
| **Agility** | 92.1 | 83.7 | 100 | 75.1 | 72.7 | 72.9 | 88.1 | 78.8 | 81.9 | 70.8 |
| **Cultural Influence** | 60.4 | 81.2 | 88.8 | 79.5 | 96.5 | 64.4 | 55 | 41.7 | 56.2 | 63.9 |
| **Entrepreneurship** | 100 | 96.9 | 99.7 | 86.4 | 68.3 | 81.7 | 81.1 | 71 | 73.2 | 81.3 |
| **Heritage** | 49.9 | 76.6 | 51.4 | 61.5 | 91 | 37.9 | 39.9 | 27.5 | 31.4 | 45 |
| **Movers** | 18.8 | 63.1 | 32.4 | 12.4 | 20.6 | 48.3 | 12.8 | 15.5 | 18 | 33 |
| **Open for Business** | 62 | 55.3 | 50.8 | 57.5 | 56.7 | 40.9 | 76.4 | 82 | 81.9 | 100 |
| **Power** | 81.6 | 63.2 | 100 | 79.5 | 63.3 | 64.7 | 43.3 | 16.9 | 22 | 26 |
| **Quality of Life** | 90 | 71.7 | 53.2 | 79.6 | 66.4 | 46.9 | 97.9 | 99.9 | 100 | 96.7 |

Note:

**Table 2.** Literature Summary on Infrastructure Portfolios Using Econometric Models.

| Studies | Model Presented | Variables | Scale/location | Data Period |
|---|---|---|---|---|
| Aschauer (1989) | Cobb-Douglas production function | Productivity, public capital | Country/U.S. | 1949-1985 annual data |
| Aschauer (1989) | Cobb-Douglas production function | Private and public capital | U.S. | 1925-1985 annual data |
| Sturm and de Haan (1995) | Cobb-Douglas production function and cointegration, OLS | Productivity, public capital | U.S.,<br>Netherlands | 1949-1985 annual data<br>1960-1990 annual data |
| Devarajan et al. (1996) | Cobb-Douglas production function | public capital | Global/43 developing countries | 1970-1990 annual data |
| Agenor et al. (2005) | VAR | GDP, public capital | Egypt, Jordan, and Tunisia | 1965-2002 annual data |
| Vu Le * and Suruga (2005) | Regression (ordinary least squares) | FDI, public capital | Global/105 developing and developed countries | 1970-2001 annual data |
| Agbelie (2014) | Regression (ordinary least squares) | GDP, PPI, Census, public capital | Global/40 countries | 1992-2010 annual data |
| Bianchi et al. (2014) | Regression (ordinary least squares), Value-at-Risk | MSCI, DJBAITRI, MGINATRI | Country/USA | 1927-2010 monthly data |
| (Bird et al. 2014) | GARCH | UBS global supply chain infrastructure and utilities index | Country/USA, Australia | 1995-2009 monthly data |
| Arbués et al. (2015) | Cobb-Douglas production function and Spatial Durbin Model | Private, Census, and public capital | Country/Spain | 1986-2006 annual data |
| Ben Ammar and Eling (2015) | Regression | RM-Rf, SMB, HML, and MOM | Country/U.S. | 1983-2011 monthly data |
| Farhadi (2015) | Generalized Method of Moments (GMM) for panel data model | Productivity, capital, domestic stock of knowledge, domestic research intensity, international knowledge, human capital, population growth drag | Global/18 OECD countries | 1870-2009 annual data |
| Achour and Belloumi (2016) | Johansen multivariate cointegration | Per capita (transportation energy consumption, value-added, infrastructure, and $CO_2$ emissions) | Country/Tunisia | 1971-2012 annual data |
| Álvarez et al. (2016) | The cobb-Douglas production function for the spatial autoregressive panel data model. | GDP, private, and public capital | Country/Spain | 1980-2007 annual data |
| Maroto and Zofío (2016) | Malmquist indices using DEA | Census and transport costs | Country/Spain | 1995-2005annual data |
| Jiang et al. (2017) | structural equations model | structural equations model, GDP, employment | Country/China | 1986-2011 annual data |
| Maparu and Mazumder (2017) | VAR | Census, infrastructure data | Country/India | 1990-2011 annual data |
| Saidi and Hammami (2017) | Generalized Method of Moments (GMM) for panel data model | World Development Indicators | Global/75 countries | 2000-2014 annual data |
| Kim et al. (2018) | fixed effect count model | Business statistics, Census data | Country/South Korea | 1997-2006 annual data |
| Okolo et al. (2018) | autoregressive distributed lag | Private and public capital | Country/Nigeria | 1970-2017 annual data |
| Shi et al. (2020) | Spatial panel data models | GDP, employment, and public capital | Country/China | 2002-2016 annual data |
| Na (2023) | VAR on Internet network Investment | Income, investment of financial statements from 16 telecom companies | Country/U.S. | 2012-2020 quarterly data |
| Umar et al. (2023) | VAR | FTSE Macquarie Global Supply Chain Infrastructure Index | Global/Europe, Australasia, North America, Japan, and Asia-Pacific | 2004-2015 monthly data |

Note:

Table 3. Descriptions of the variables in this study

| Variable | Symbol | Description | Source |
|---|---|---|---|
| Portfolios on global supply chain infrastructure | GII, GLFOX, FGIAX, CSUAX | Investment funds for global supply chain infrastructure | CRSP and MSCI |
| Energy Market | WTI | West Texas Intermediate (WTI) - Cushing, Oklahoma, Dollars per Barrel, Daily, Not Seasonally Adjusted | Federal Reserve Bank |
| Investor sentiment | VIX | Volatility index to measure volatility in the stock market accompanied by market fear | CRSP |
| Global shipping costs | BDI | Baltic Exchange Dry Index | Bloomberg |

Note:

Table 4. ESG risk score of four portfolios on global supply chain infrastructure investment

| Ticker | ESG Risk Score | Environment Risk Score | Social Risk Score | Governance Risk Score |
|---|---|---|---|---|
| GII | 24 | 2.0 | 12.7 | 9.6 |
| GLFOX | 22 | 6.0 | 9.1 | 5.7 |
| FGIAX | 23 | 7.8 | 8.0 | 5.3 |
| CSUAX | 25 | 9.6 | 8.5 | 5.2 |

Note:

Table 5. Descriptive statistics of four portfolios, VIX, WTI, and BDI

| Variable | Mean | Median | SD | Skewness | Kurtosis | J-B test | Q2(20) |
|---|---|---|---|---|---|---|---|
| **GII** | 0.000176 | 0.00074 | 0.013845 | -2.054*** | 29.176*** | 122695.290*** | -10.501*** |
| **GLFOX** | 0.000634 | 0 | 0.018063 | -1.464*** | 24.344*** | 84968.326*** | -22.450*** |
| **FGIAX** | 0.000367 | 0 | 0.016418 | -2.259*** | 34.850*** | 174533.623*** | -22.635*** |
| **CSUAX** | 0.000316 | 0.001051 | 0.015209 | -1.659*** | 27.794*** | 110737.223*** | -15.537*** |
| **BDI** | -0.00027 | -0.00025 | 0.008206 | 0.203*** | 9.367*** | 12424.219*** | -3.058*** |
| **WTI** | -0.00023 | 0.000486 | 0.012639 | -0.565*** | 6.177*** | 5572.940*** | -16.731*** |
| **VIX** | -0.00029 | -2.71E-06 | 0.024045 | -1.824*** | 151.441*** | 3243285.099*** | -28.021*** |

Note: (. p-value<=0.1, * p-value<=0.05, ** p-value<=0.01, ***, p-value<=0.005). The Ljung-Box test indicates that the null hypothesis of no autocorrelation up to order 20 for the squared standard residuals (Q2(20)) cannot be rejected at the 1% significance level.

Table 6. ADF test results for variables and their value at first-order difference

| Ticker | ADF test at first difference (Lag) | p-value |
|---|---|---|
| **GII** | -11.7***(29) | 1.58E-21 |
| **GLFOX** | -11.52***(29) | 4.12E-21 |
| **FGIAX** | -12.16***(29) | 1.52E-22 |
| **CSUAX** | -12.47***(29) | 3.23E-23 |
| **BDI** | -12.59***(29) | 1.87E-23 |
| **WTI** | -59.61***(0) | 0 |
| **VIX** | -13.83***(27) | 7.75E-26 |

Note: (.p-value<=0.1, * p-value<=0.05, ** p-value<=0.01, ***, p-value<=0.005)

Table 7. VAR conditional correlation results

| | GII | GLFOX | FGIAX | CSUAX | BDI | WTI | VIX |
|---|---|---|---|---|---|---|---|
| **GII** | 1 | 0.81 | 0.93 | 0.91 | -0.01 | 0.33 | -0.25 |
| **GLFOX** | | 1 | 0.83 | 0.82 | -0.01 | 0.2 | -0.16 |
| **FGIAX** | | | 1 | 0.95 | -0.02 | 0.3 | -0.23 |
| **CSUAX** | | | | 1 | -0.01 | 0.28 | -0.22 |
| **BDI** | | | | | 1 | -0.03 | 0.07 |
| **WTI** | | | | | | 1 | -0.25 |
| **VIX** | | | | | | | 1 |

Note:

Table 8. VAR partial correlation results

| | GII | GLFOX | FGIAX | CSUAX | BDI | WTI | VIX |
|---|---|---|---|---|---|---|---|
| **GII** | 1 | 0.16 | 0.44 | 0.2 | 0.03 | 0.15 | -0.09 |
| **GLFOX** | | 1 | 0.2 | 0.12 | 0 | -0.1 | 0.06 |
| **FGIAX** | | | 1 | 0.65 | -0.01 | 0.03 | -0.03 |
| **CSUAX** | | | | 1 | 0 | -0.02 | 0.03 |
| **BDI** | | | | | 1 | -0.01 | 0.07 |
| **WTI** | | | | | | 1 | -0.17 |
| **VIX** | | | | | | | 1 |

Note:

Table 9. CHOW test on structural stability of VAR model

| | Test statistic | 95% critical value | p-value |
|---|---|---|---|
| **Break-point Test** | 35475.2 | 27797.5 | <2e-16 *** |
| **Sample-split test** | 399.7 | 386.2 | 0.008 ** |

Note: (. p-value<=0.1, * p-value<=0.05, ** p-value<=0.01, ***, p-value<=0.005)

Table 10. Cointegration t-test results of variables

| | GII | GLFOX | FGIAX | CSUAX | BDI | WTI | VIX |
|---|---|---|---|---|---|---|---|
| **GII** | | -65.94*** | -34.16*** | -27.33*** | -16.85*** | -25.26*** | -31.81*** |
| **GLFOX** | -25.37*** | | -25.19*** | -30.51*** | -16.86*** | -25.26*** | -31.6*** |
| **FGIAX** | -36.16*** | -62.58*** | | -28.12*** | -16.86*** | -25.23*** | -31.79*** |
| **CSUAX** | -26.17*** | -43.59*** | -26.74*** | | -16.86*** | -25.07*** | -32.23*** |
| **BDI** | -24.5*** | -24.16*** | -24.55*** | -25.2*** | | -25.17*** | -27.08*** |
| **WTI** | -24.42*** | -24.19*** | -24.43*** | -24.98*** | -16.84*** | | -26.97*** |
| **VIX** | -24.91*** | -24.63*** | -24.94*** | -26.16*** | -16.85*** | -25.08*** | |

Note: (. *p*-value<=0.1, * *p*-value<=0.05, ** *p*-value<=0.01, ***, *p*-value<=0.005)

Table 11. VAR connectedness analysis with time on variables

| | GII | GLFOX | FGIAX | CSUAX | BDI | WTI | VIX | Receiver |
|---|---|---|---|---|---|---|---|---|
| **GII** | 28.52 | 18.56 | 24.61 | 23.47 | 0.01 | 3.01 | 1.82 | 71.48 |
| **GLFOX** | 21.31 | 32.24 | 22.56 | 21.78 | 0.01 | 1.29 | 0.81 | 67.76 |
| **FGIAX** | 24.13 | 19.17 | 27.72 | 25.06 | 0.02 | 2.41 | 1.5 | 72.28 |
| **CSUAX** | 23.62 | 19.04 | 25.74 | 28.1 | 0.02 | 2.18 | 1.32 | 71.9 |
| **BDI** | 0.03 | 0.01 | 0.03 | 0.02 | 99.08 | 0.07 | 0.76 | 0.92 |
| **WTI** | 7.79 | 3.05 | 6.47 | 5.76 | 0.06 | 72.22 | 4.65 | 27.78 |
| **VIX** | 5.03 | 1.96 | 4.28 | 3.71 | 0.51 | 4.99 | 79.51 | 20.49 |
| **Giver** | 81.91 | 61.78 | 83.7 | 79.79 | 0.63 | 13.94 | 10.86 | 332.61 |
| **Inc.Own** | 110.43 | 94.02 | 111.42 | 107.89 | 99.71 | 86.17 | 90.37 | **TCI** |
| **NET** | 10.43 | -5.98 | 11.42 | 7.89 | -0.29 | -13.83 | -9.63 | **47.52** |
| **NPT** | 5 | 3 | 6 | 4 | 0 | 2 | 1 | |

Note: Givers: FGIAX, GII, CSUAX
Receivers: WTI, VIX, GLFOX, BDI

Table 12. TVP-VAR connectedness analysis on variables

| | GII | GLFOX | FGIAX | CSUAX | BDI | WTI | VIX | Receiver |
|---|---|---|---|---|---|---|---|---|
| **GII** | 27.83 | 15.28 | 21.9 | 20.93 | 0.83 | 4.62 | 8.61 | 72.17 |
| **GLFOX** | 17.41 | 32.22 | 18.46 | 18.56 | 0.78 | 3.37 | 9.19 | 67.78 |
| **FGIAX** | 21.01 | 15.43 | 26.72 | 22.32 | 0.8 | 4.4 | 9.33 | 73.28 |
| **CSUAX** | 20.46 | 15.85 | 22.79 | 27.07 | 0.82 | 4 | 9.01 | 72.93 |
| **BDI** | 2.65 | 2.65 | 3.13 | 3.07 | 84.01 | 2.11 | 2.39 | 15.99 |
| **WTI** | 8.93 | 5.91 | 9.03 | 7.92 | 1.73 | 62.08 | 4.4 | 37.92 |
| **VIX** | 12.96 | 13 | 14.93 | 14.65 | 1.85 | 3.8 | 38.81 | 61.19 |
| **Giver** | 83.41 | 68.12 | 90.23 | 87.46 | 6.81 | 22.29 | 42.92 | 401.25 |
| **Inc.Own** | 111.25 | 100.35 | 116.95 | 114.53 | 90.82 | 84.37 | 81.74 | **TCI** |
| **NET** | 11.25 | 0.35 | 16.95 | 14.53 | -9.18 | -15.63 | -18.26 | **57.32** |
| **NPT** | 4 | 3 | 6 | 5 | 0 | 1 | 2 | |

Note: Givers: FGIAX, CSUAX, GII, GLFOX
Receivers: VIX, WTI, BDI

Table 13. Net contributions across rolling windows

| Rolling Windows | GII | GLFOX | FGIAX | CSUAX | BDI | WTI | VIX |
|---|---|---|---|---|---|---|---|
| **60** | 11.03 | 0.93 | 16.68 | 14.09 | -9 | -15.37 | -18.37 |
| **120** | 11.15 | 0.51 | 16.84 | 14.29 | -9.14 | -15.43 | -18.22 |
| **180** | 11.17 | 0.46 | 16.82 | 14.27 | -9.06 | -15.4 | -18.27 |
| **360** | 11.21 | 0.38 | 16.7 | 14.1 | -8.85 | -15.27 | -18.28 |

Note:

Table 14. Comparison of hedge ratios and hedging effectiveness for 2013-2023

| Portfolio/Factor pairs | HR | HE | Factor/Portfolio pairs | HR | HE |
|---|---|---|---|---|---|
| Global supply chain infrastructure portfolios as a long-position | | | Global supply chain infrastructure portfolios as short position | | |
| GII/BDI | 0.07 | -0.01 | BDI/GII | -0.01 | 0.01 |
| GII/WTI | 0.43 | 0.17 | BDI/GLFOX | -0.01 | 0.01 |
| GII/VIX | -103.04 | 0.35 | BDI/FGIAX | -0.02 | 0.01 |
| GLFOX/BDI | 0.1 | -0.01 | BDI/CSUAX | -0.01 | 0.01 |
| GLFOX/WTI | 0.4 | 0.11 | WTI/GII | 0.41 | 0.19 |
| GLFOX/VIX | -121.41 | 0.32 | WTI/GLFOX | 0.29 | 0.12 |
| FGIAX/BDI | 0.06 | 0 | WTI/FGIAX | 0.38 | 0.2 |
| FGIAX/WTI | 0.47 | 0.14 | WTI/CSUAX | 0.4 | 0.19 |
| FGIAX/VIX | -139.7 | 0.37 | VIX/GII | -0.67 | 0.42 |
| CSUAX/BDI | 0.07 | 0 | VIX/GLFOX | -0.69 | 0.37 |
| CSUAX/WTI | 0.41 | 0.13 | VIX/FGIAX | -0.69 | 0.48 |
| CSUAX/VIX | -121.91 | 0.35 | VIX/CSUAX | -0.73 | 0.44 |

Note:

Table 15. Hedge ability of hedge ratios and hedging effectiveness value for portfolios before and during COVID-19

| Portfolio/Factor pairs | before COVID-19 | | after COVID-19 | | Factor/Portfolio pairs | before COVID-19 | | after COVID-19 | |
|---|---|---|---|---|---|---|---|---|---|
| | HR | HE | HR | HE | | HR | HE | HR | HE |
| Global supply chain infrastructure portfolios as a long position | | | | | Global supply chain infrastructure portfolios as short position | | | | |
| GII/BDI | 0.03 | 0 | 0.12 | 0 | BDI/GII | -0.01 | 0.01 | 0 | 0 |
| GII/WTI | 0.3 | 0.18 | 0.88 | 0.16 | BDI/GLFOX | -0.02 | 0.01 | 0 | 0 |
| GII/VIX | -5.7 | 0.33 | -300.42 | 0.31 | BDI/FGIAX | -0.02 | 0.02 | 0 | 0 |
| GLFOX/BDI | 0.05 | 0 | 0.2 | -0.01 | BDI/CSUAX | -0.02 | 0.02 | 0 | 0 |
| GLFOX/WTI | 0.24 | 0.08 | 0.94 | 0.13 | WTI/GII | 0.51 | 0.2 | 0.17 | 0.1 |
| GLFOX/VIX | -8.42 | 0.3 | -336.45 | 0.23 | WTI/GLFOX | 0.39 | 0.13 | 0.08 | 0.07 |
| FGIAX/BDI | 0.02 | 0 | 0.12 | 0 | WTI/FGIAX | 0.5 | 0.22 | 0.11 | 0.09 |
| FGIAX/WTI | 0.32 | 0.18 | 1.01 | 0.14 | WTI/CSUAX | 0.53 | 0.21 | 0.11 | 0.08 |
| FGIAX/VIX | -6.72 | 0.38 | -409.39 | 0.33 | VIX/GII | -0.95 | 0.44 | 0 | 0.27 |
| CSUAX/BDI | 0.03 | 0 | 0.01 | 0 | VIX/GLFOX | -1.05 | 0.43 | 0 | 0.45 |
| CSUAX/WTI | 0.28 | 0.17 | 0.9 | 0.12 | VIX/FGIAX | -1.03 | 0.55 | 0 | 0.19 |
| CSUAX/VIX | -6.01 | 0.37 | -379.15 | 0.33 | VIX/CSUAX | -1.1 | 0.5 | 0 | 0.06 |

Note:

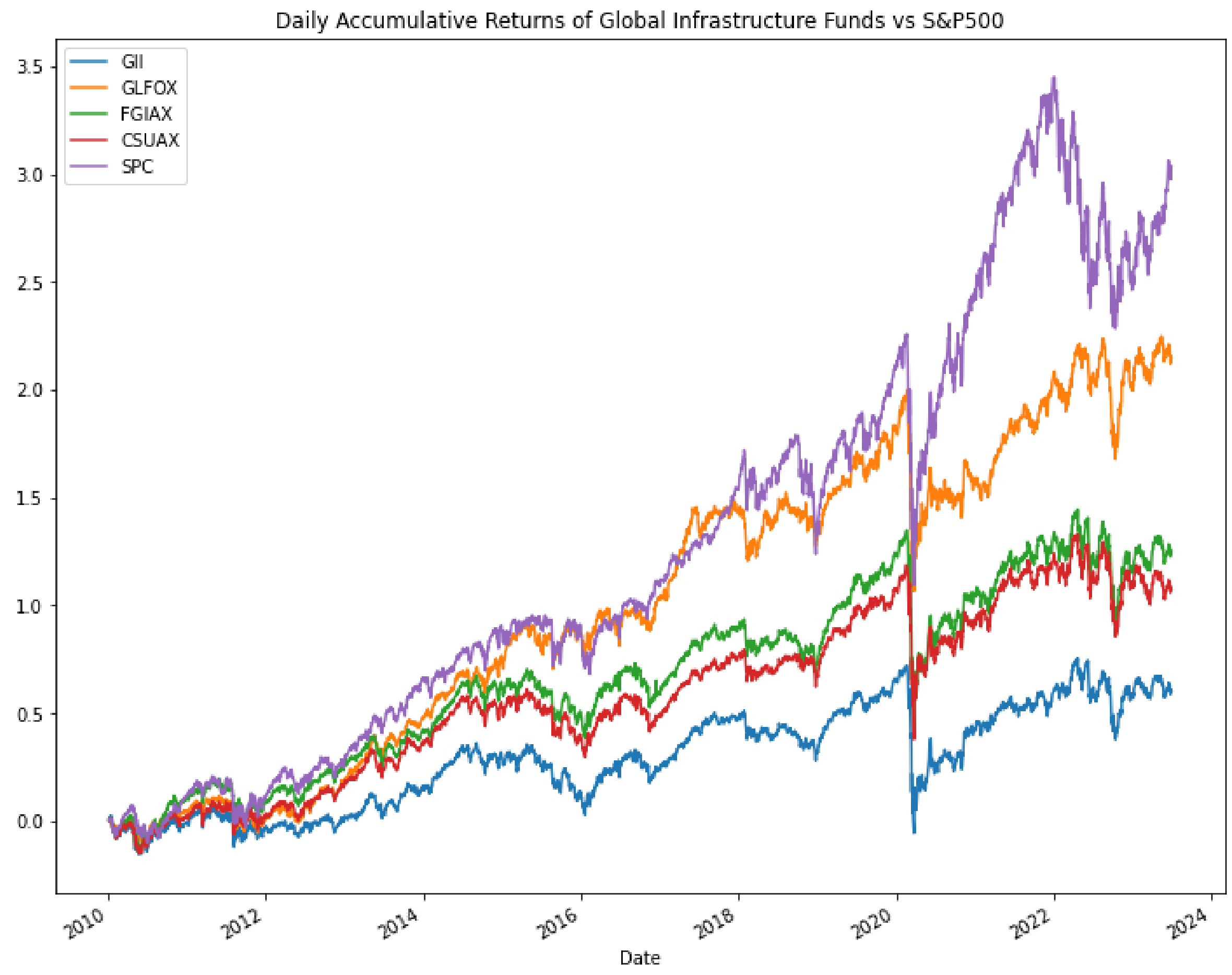

Figure 1. Daily cumulative return of portfolios for global supply chain infrastructure

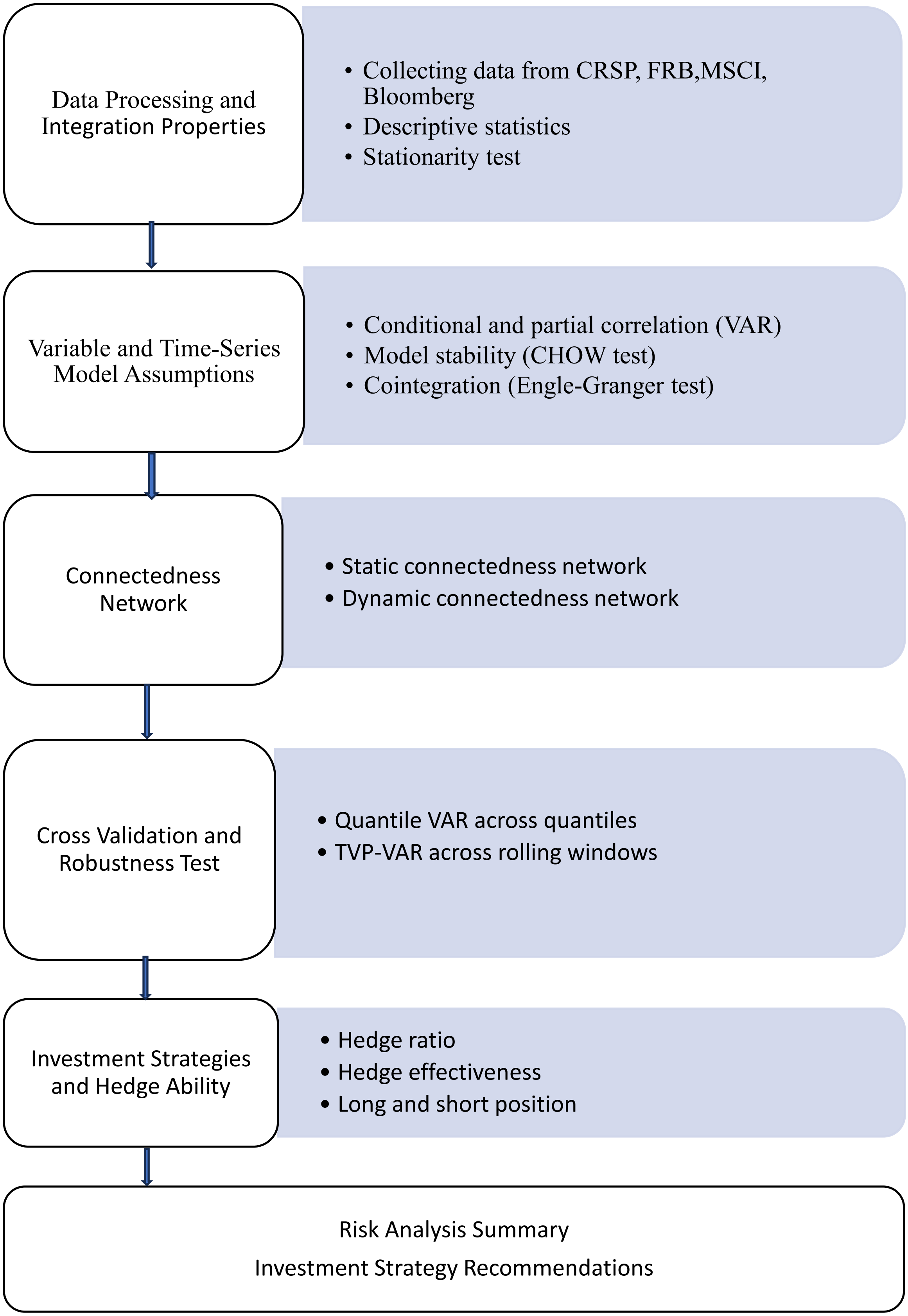
Data Processing and Integration Properties
• Collecting data from CRSP, FRB,MSCI, Bloomberg
• Descriptive statistics
• Stationarity test
Variable and Time-Series Model Assumptions
• Conditional and partial correlation (VAR)
• Model stability (CHOW test)
• Cointegration (Engle-Granger test)
Connectedness Network
• Static connectedness network
• Dynamic connectedness network
Cross Validation and Robustness Test
• Quantile VAR across quantiles
• TVP-VAR across rolling windows
Investment Strategies and Hedge Ability
• Hedge ratio
• Hedge effectiveness
• Long and short position
Risk Analysis Summary
Investment Strategy Recommendations

Figure 2. Analytics modules and workflow in the framework

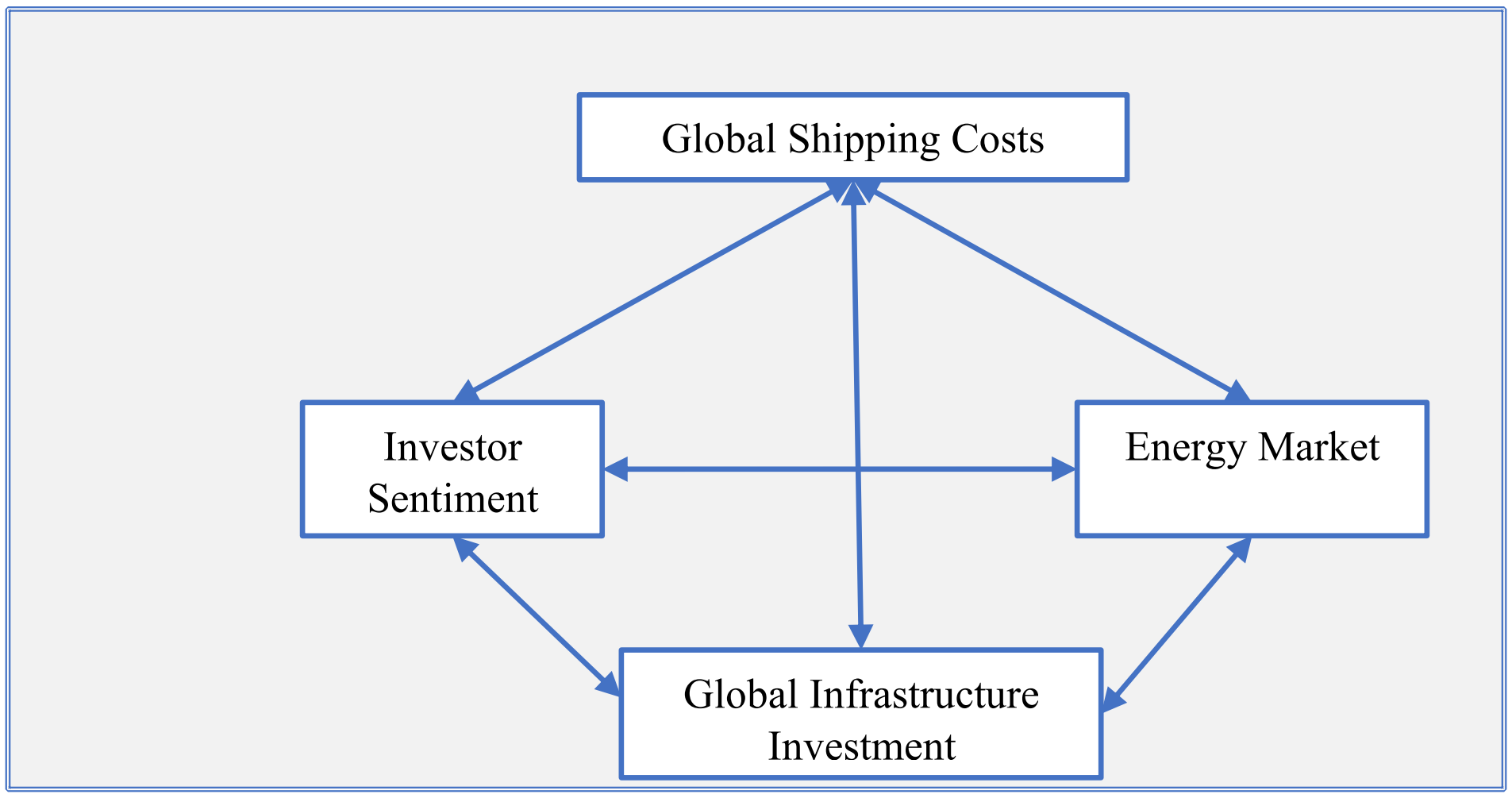

Figure 3. Cointegration relationship of variables

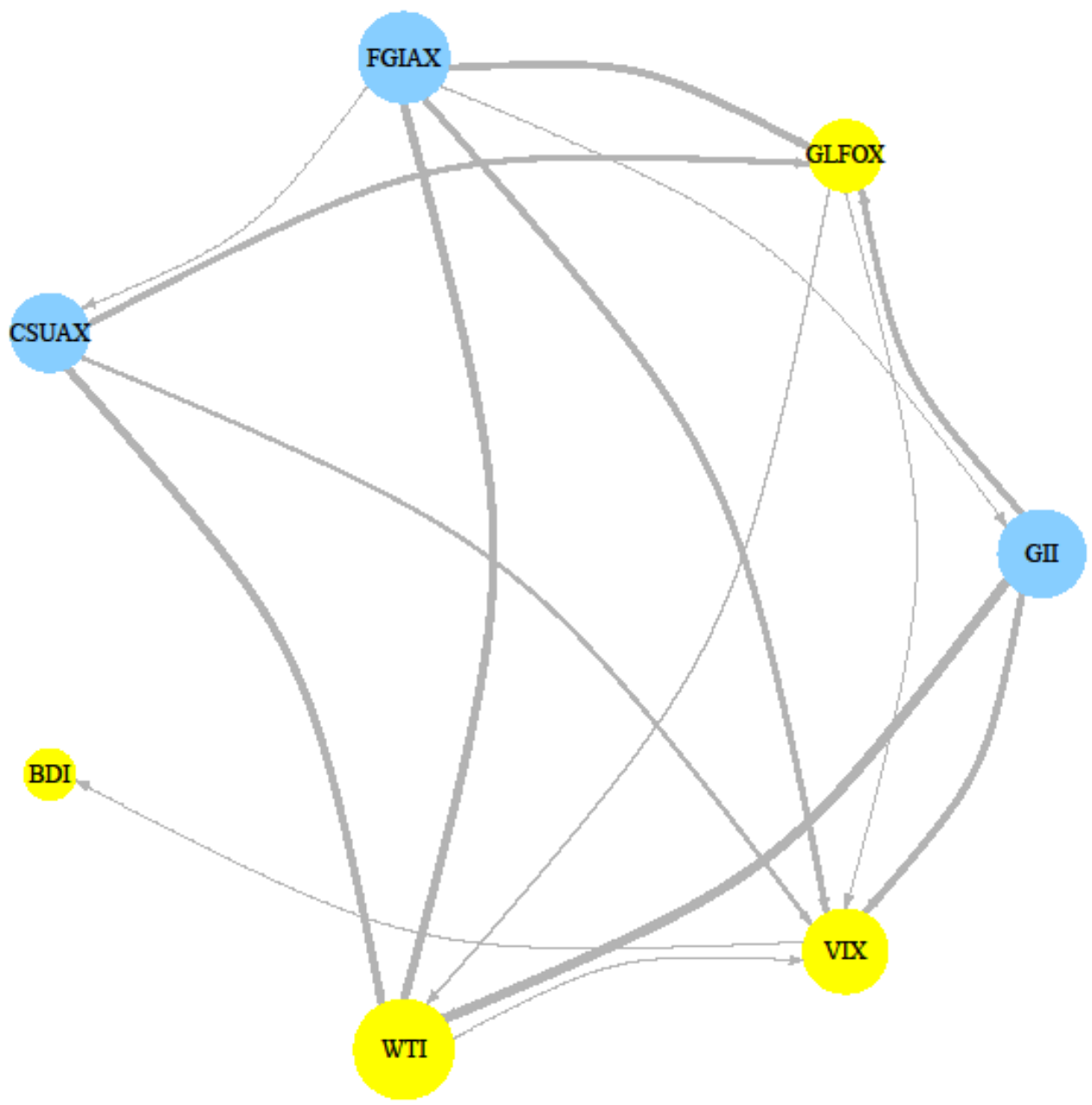

Figure 4. NPDC measures the plot of VAR connectedness with time

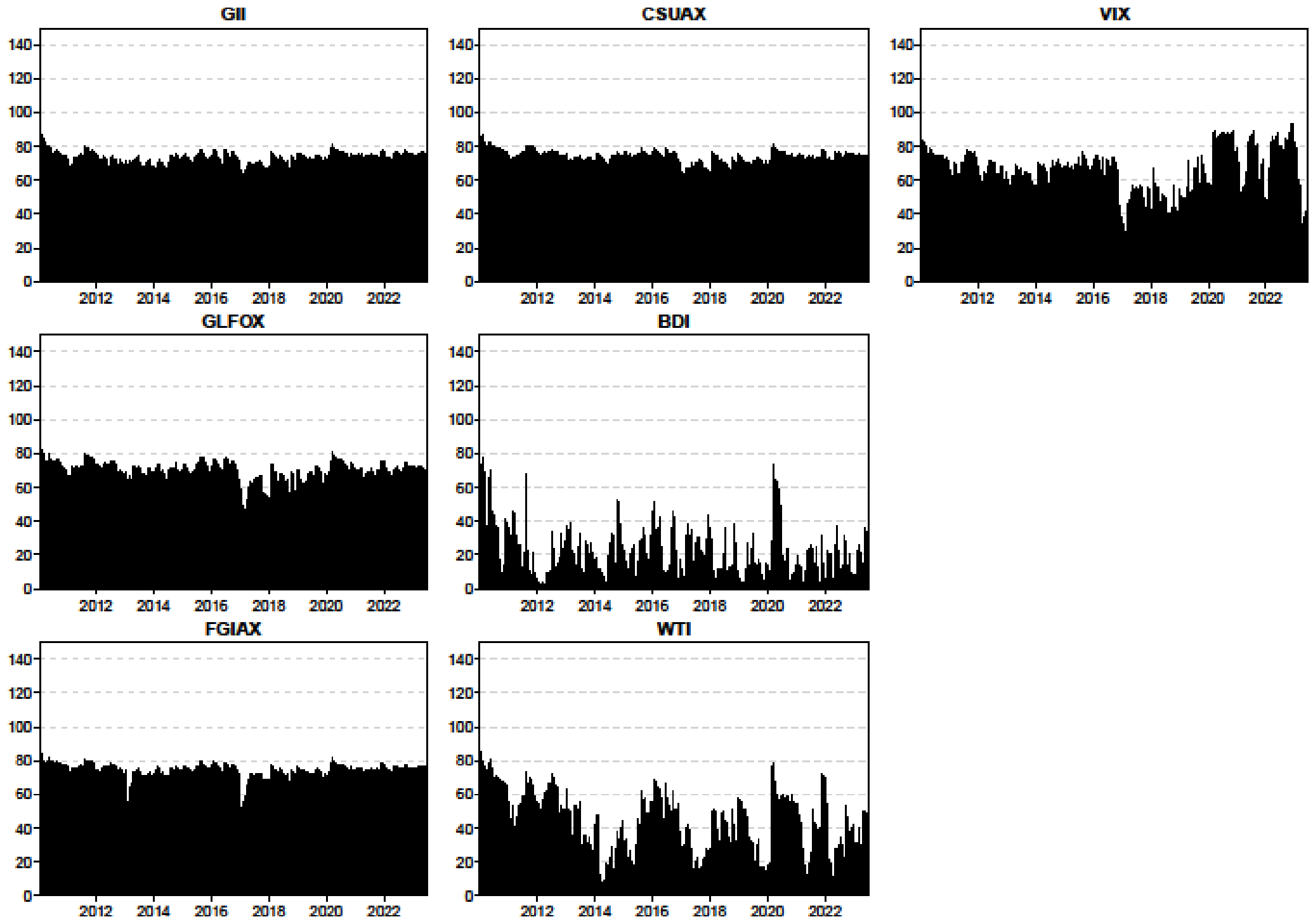

Figure 5. Variable as the spillover receiver using TVP-VAR connectedness analysis

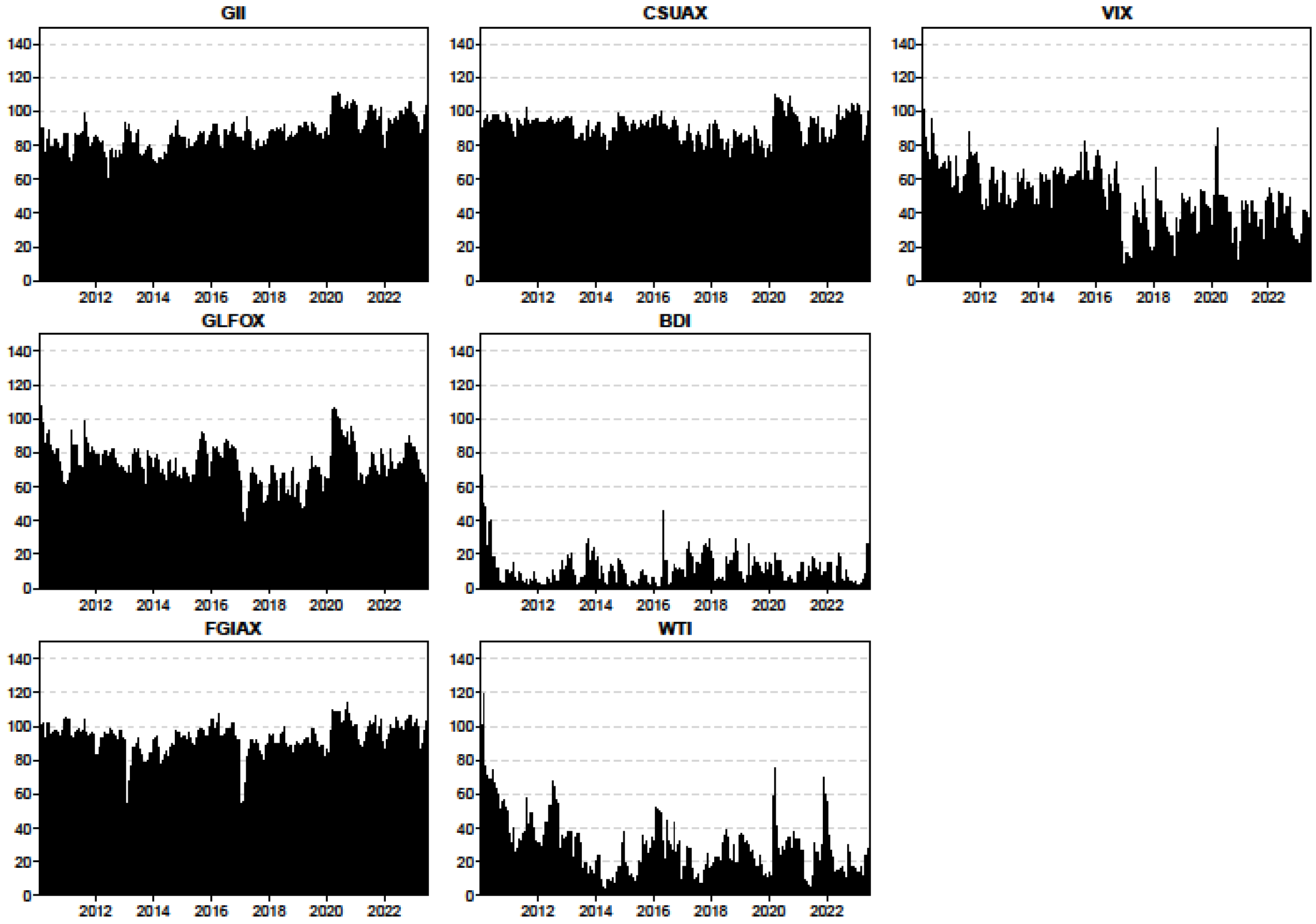

Figure 6. Variable as the spillover giver using TVP-VAR connectedness analysis

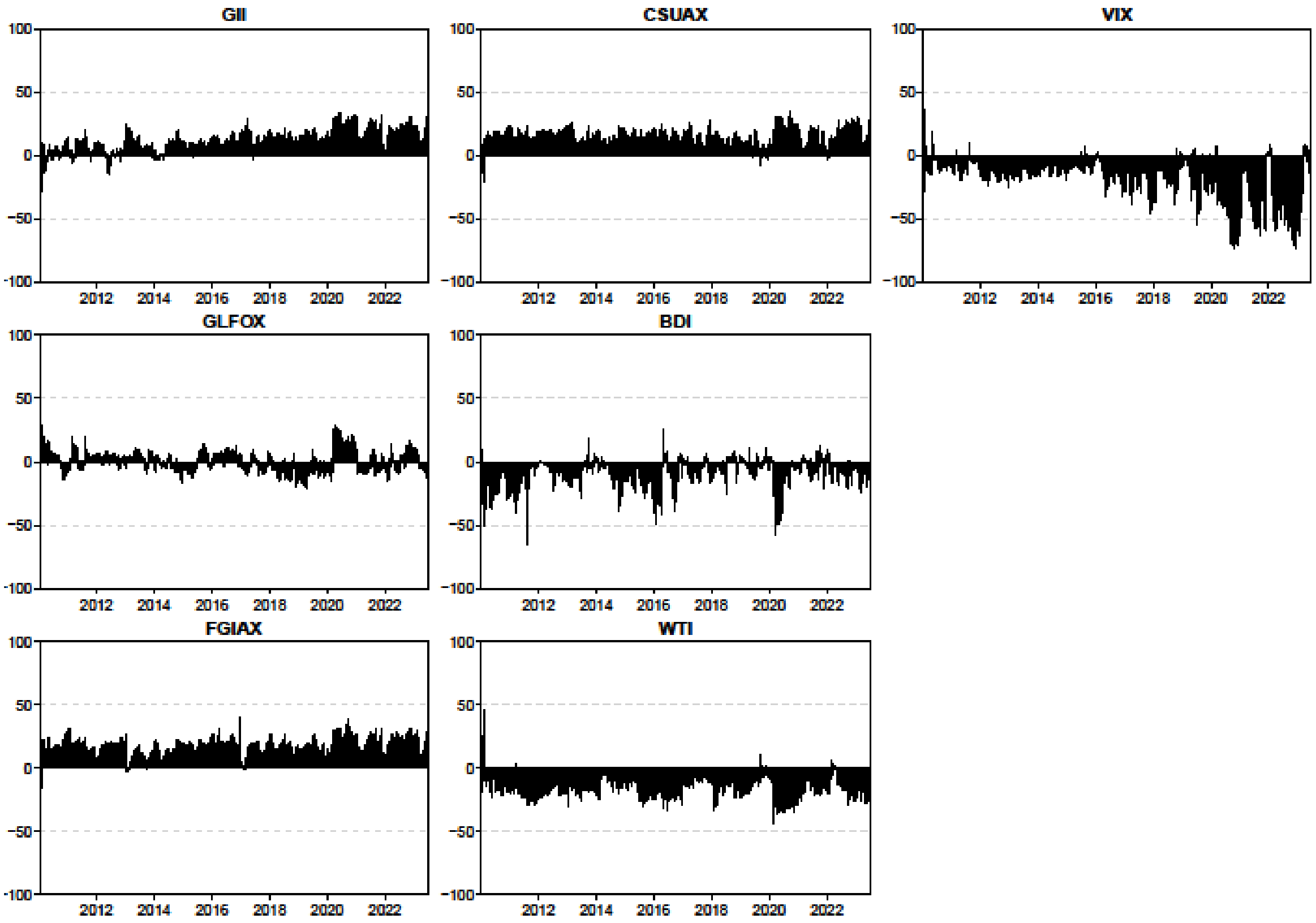

Figure 7. Variable's NET spillover using TVP-VAR connectedness analysis

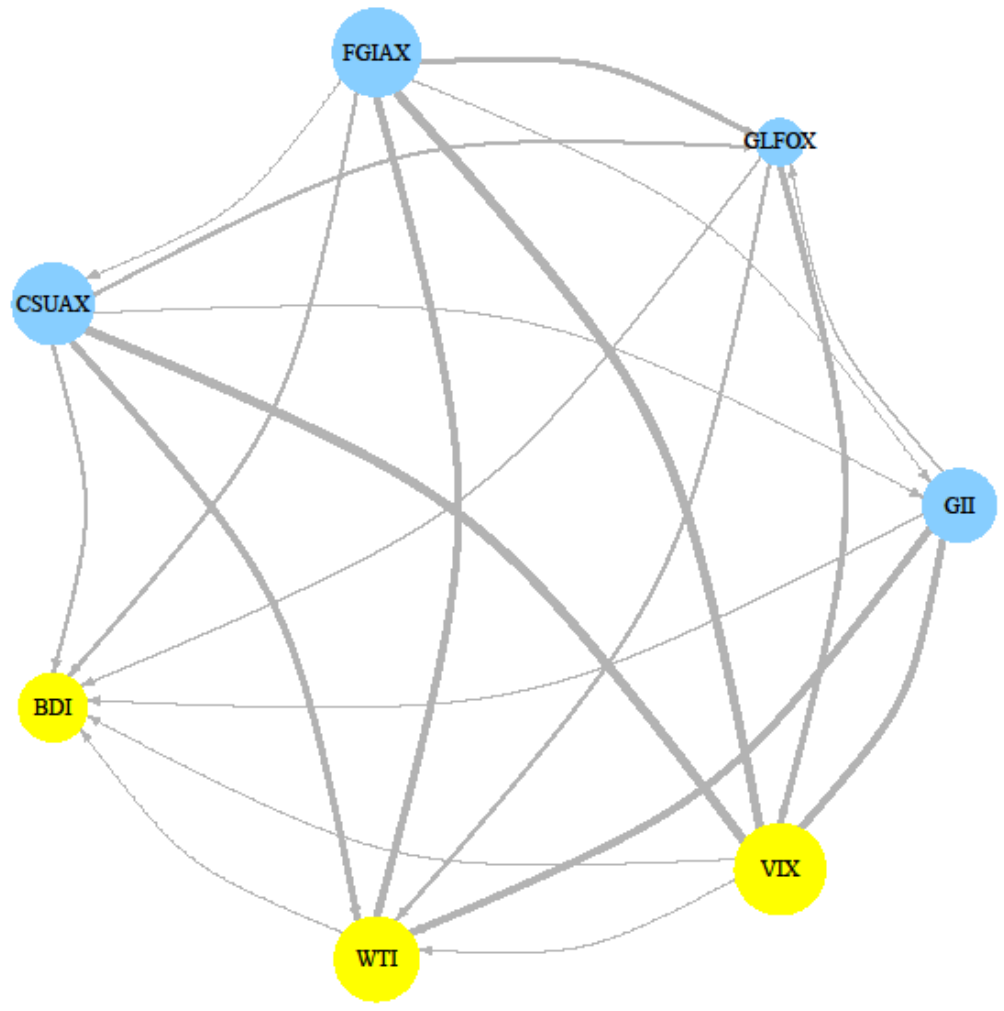

Figure 8. NPDC measure plot of TVP-VAR connectedness analysis

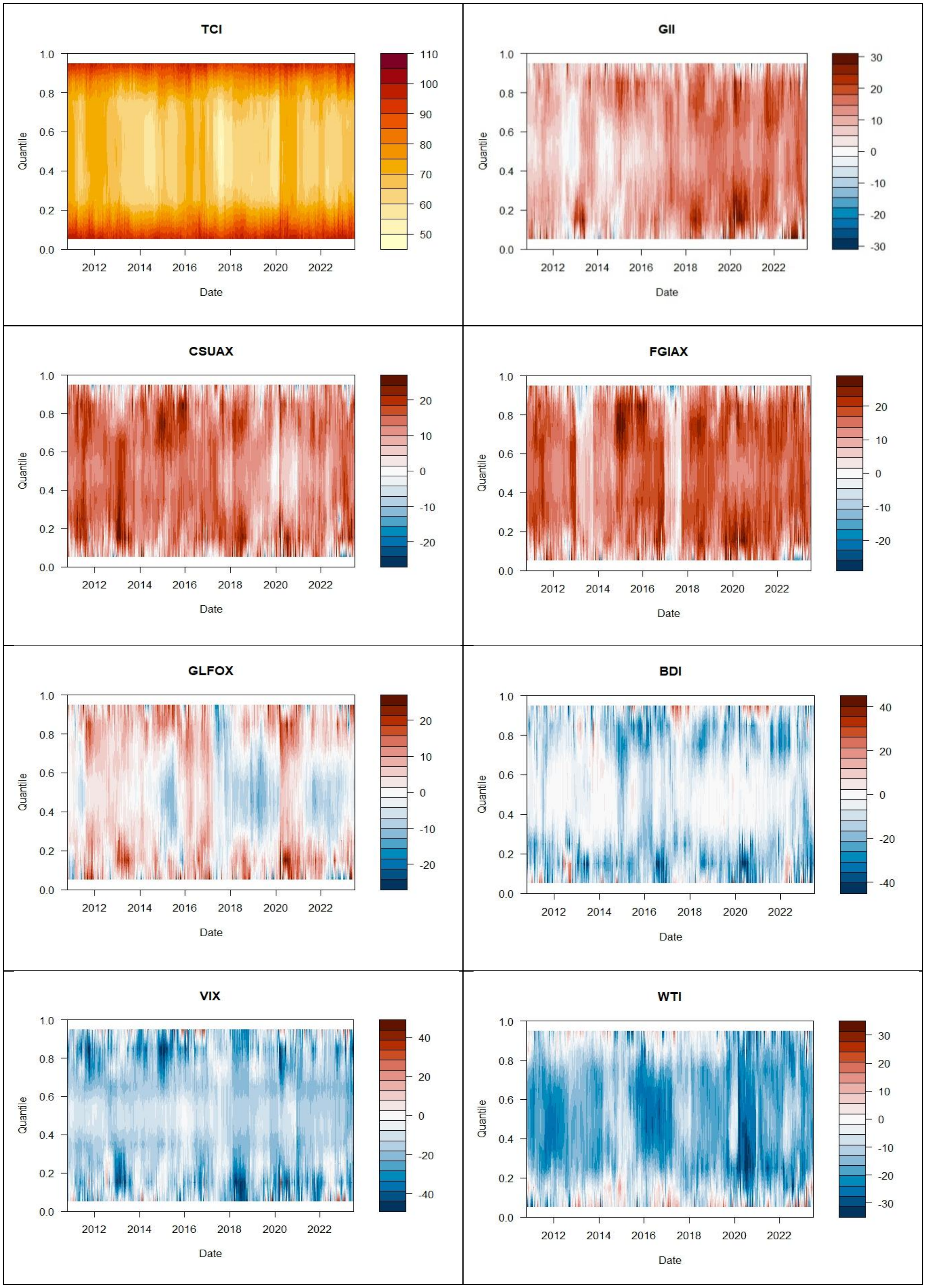

Figure 9. Heatmaps of QVAR on all variables across the quantiles

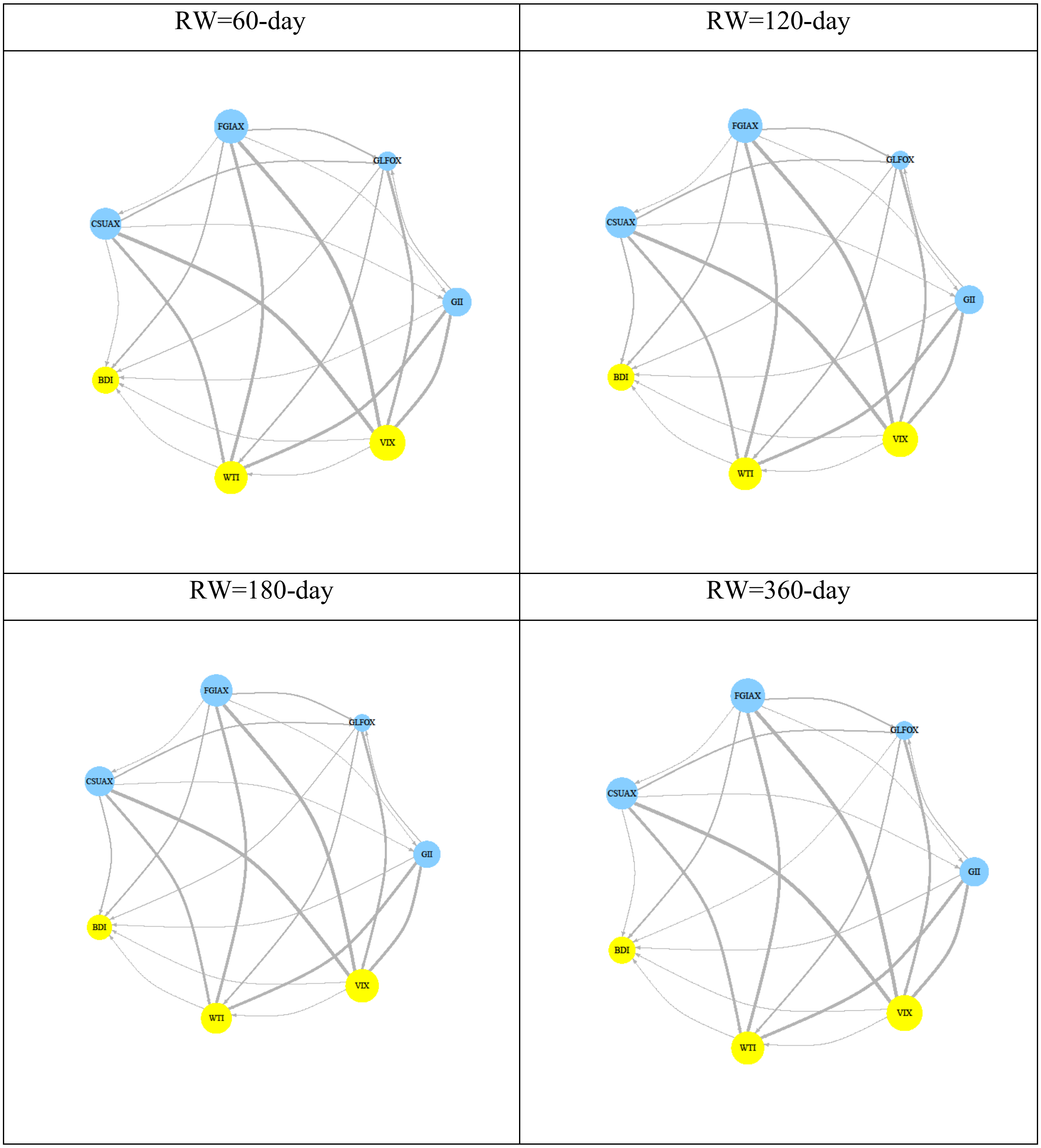

Figure 10. Dynamic Connectedness Across Rolling Windows (RW)

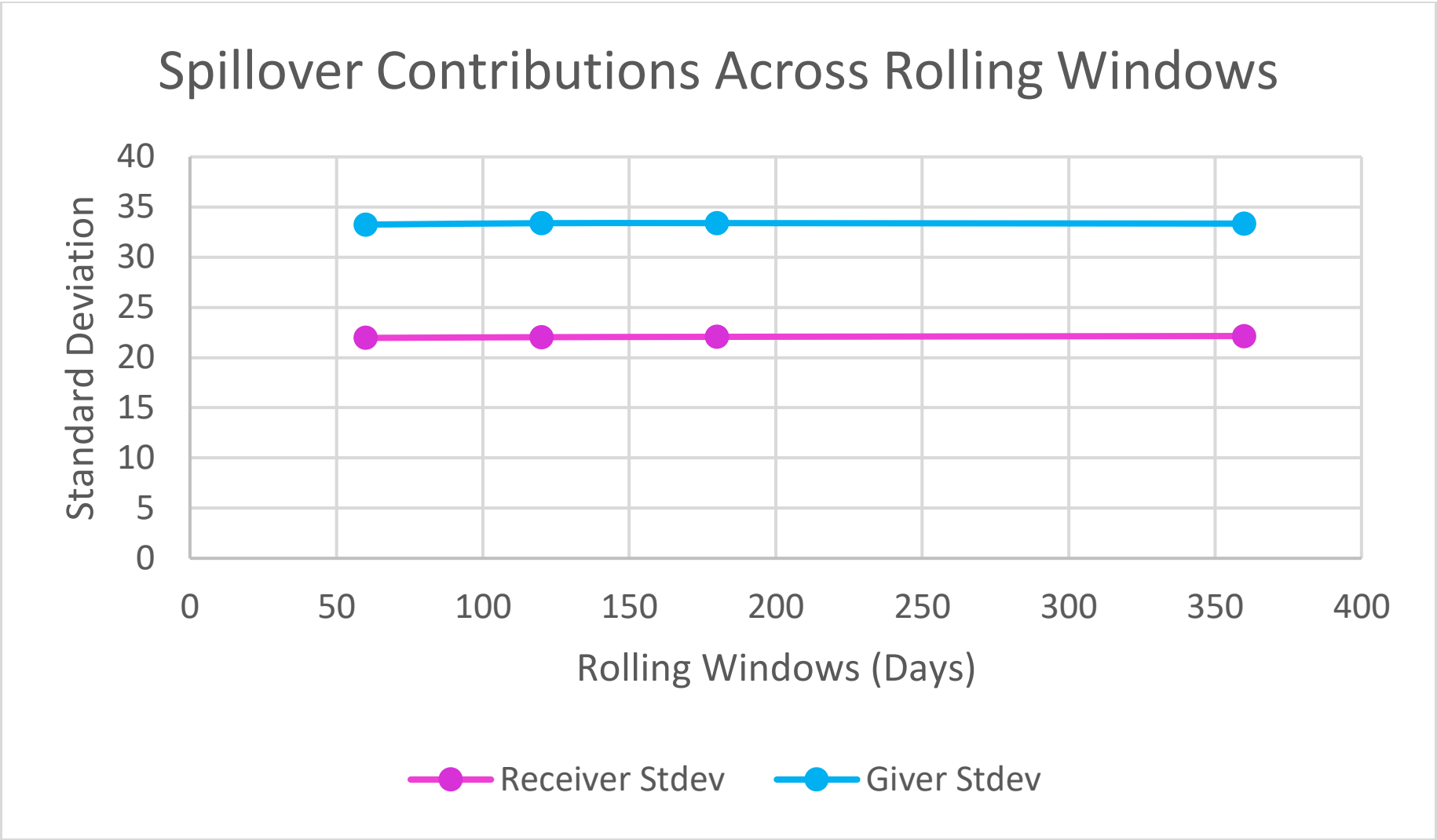

Figure 11. Spillover contributions of receivers and givers across rolling windows

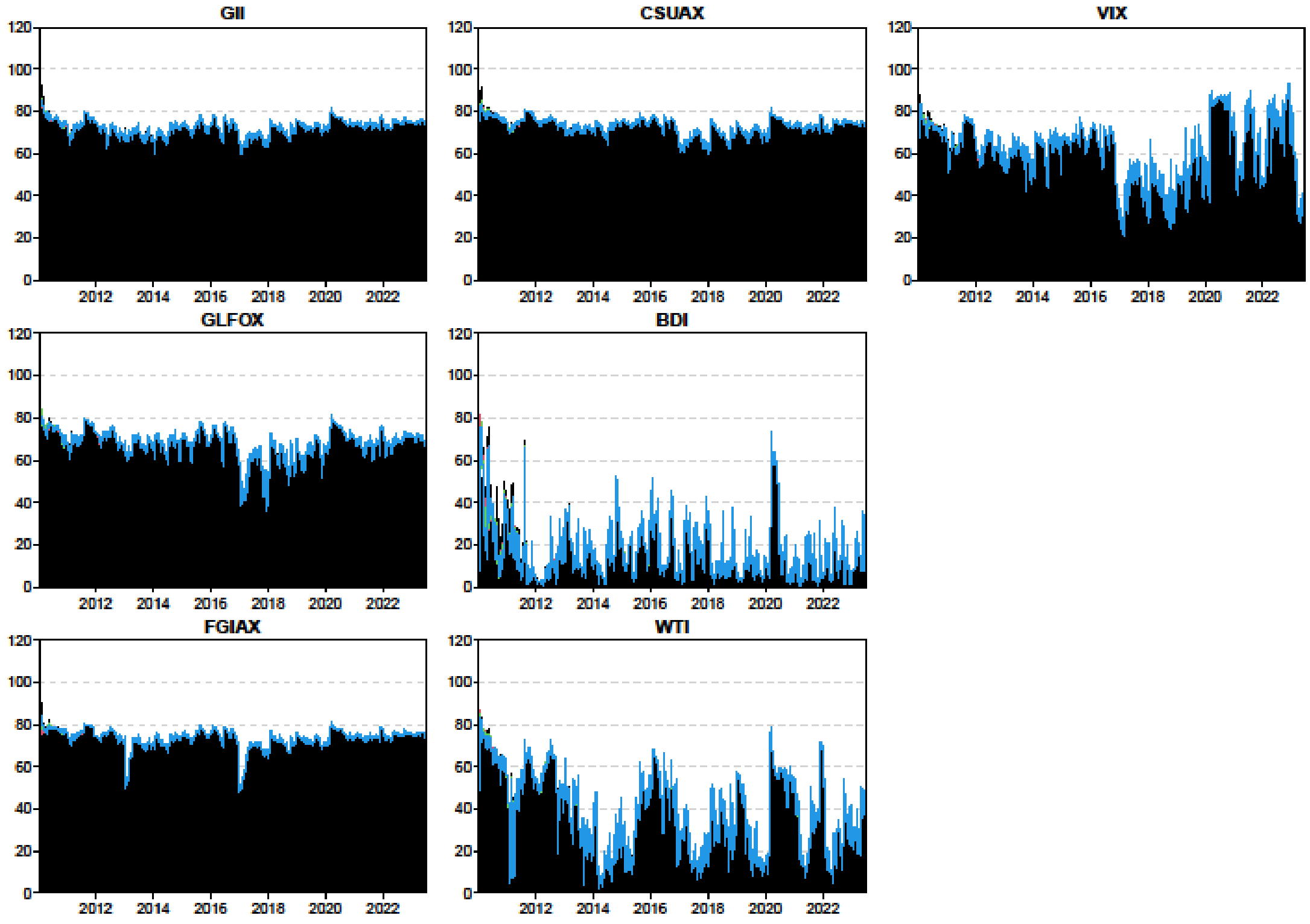

Figure 12. Variable as the spillover receiver across rolling windows

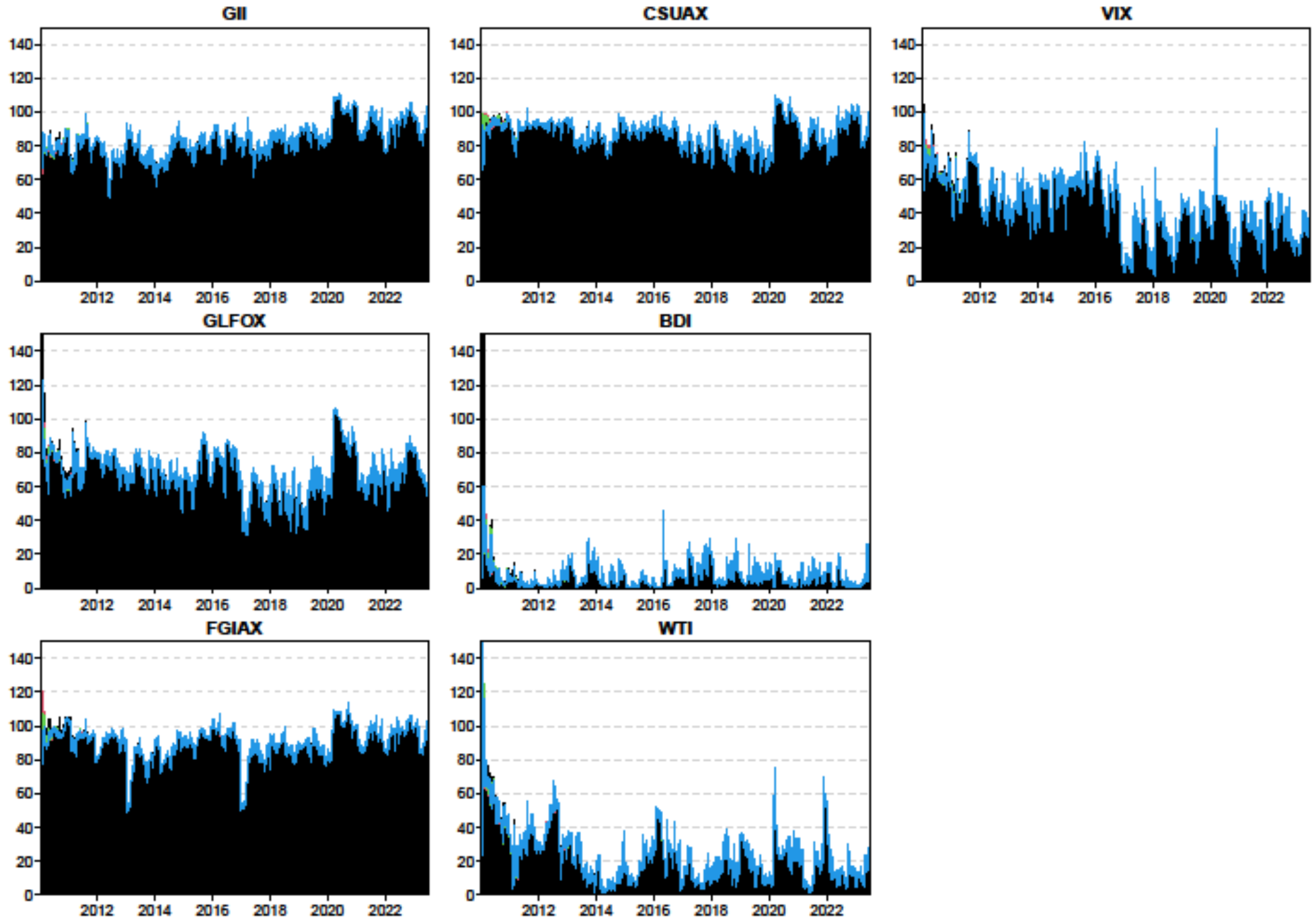

Figure 13. Variable as the spillover giver across rolling windows

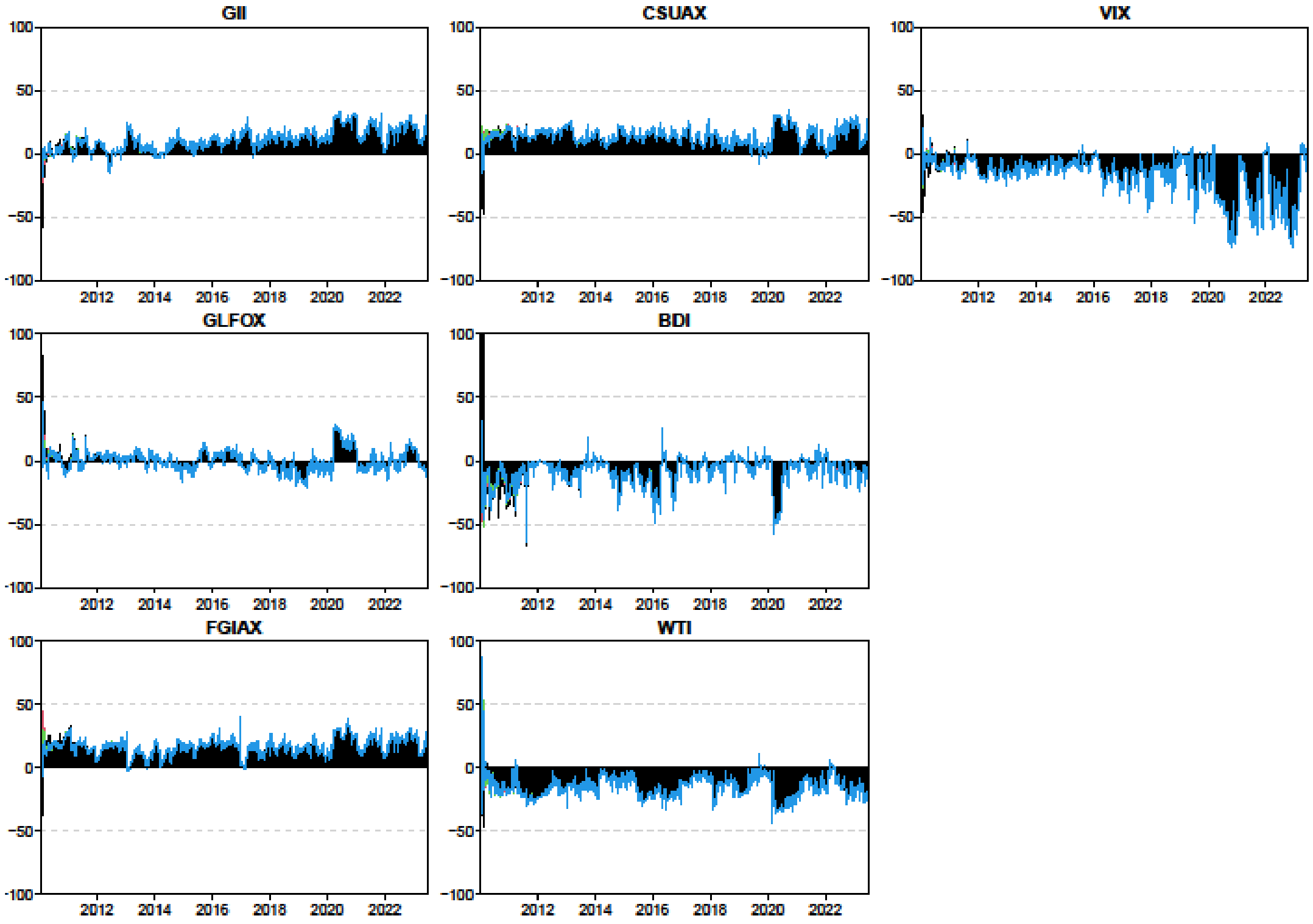

Figure 14. Variable's NET spillover across rolling windows

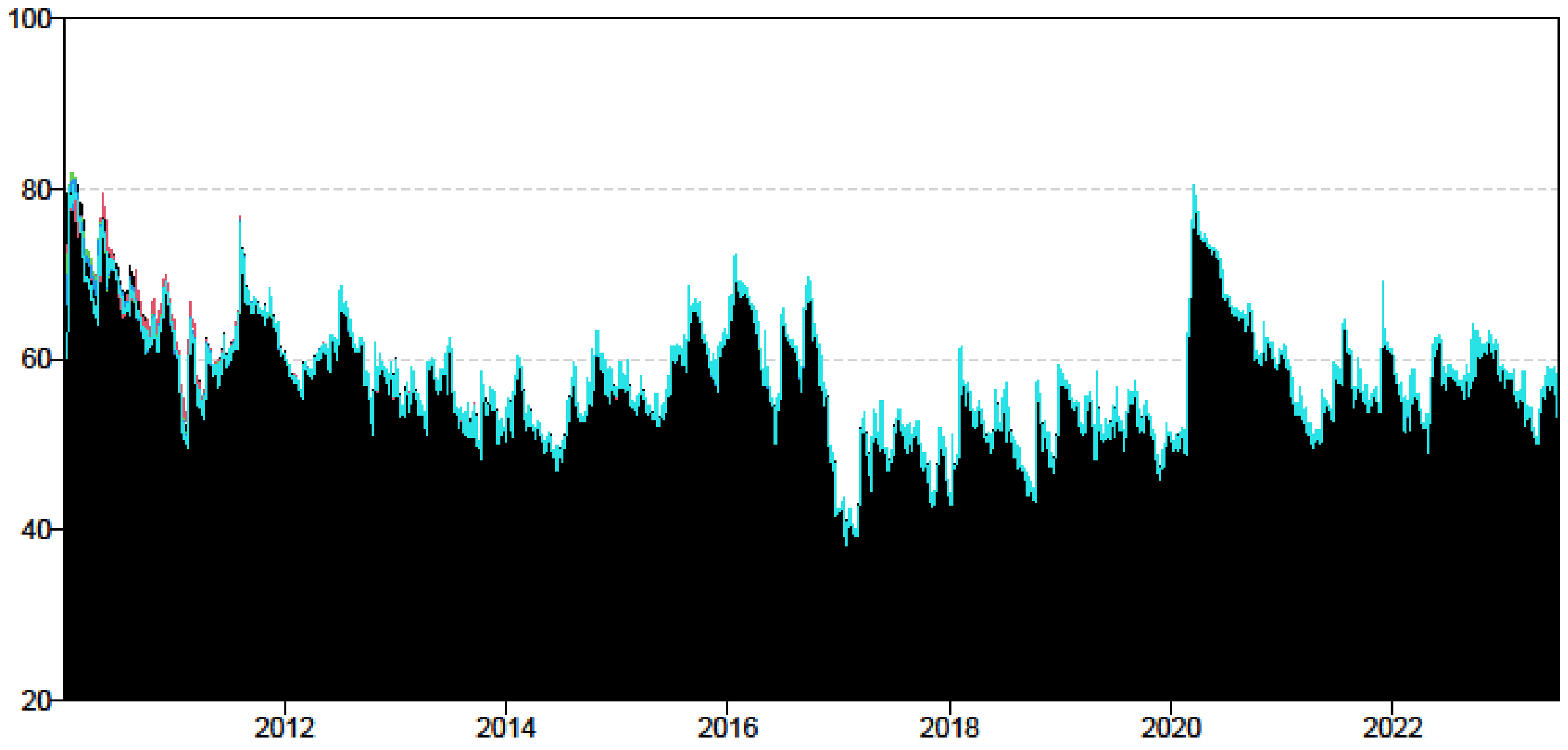

Figure 15. TCI of spillover plot across rolling windows of 60 (red color), 120 (green color), 180 (blue color), 360 (aquamarine color) days

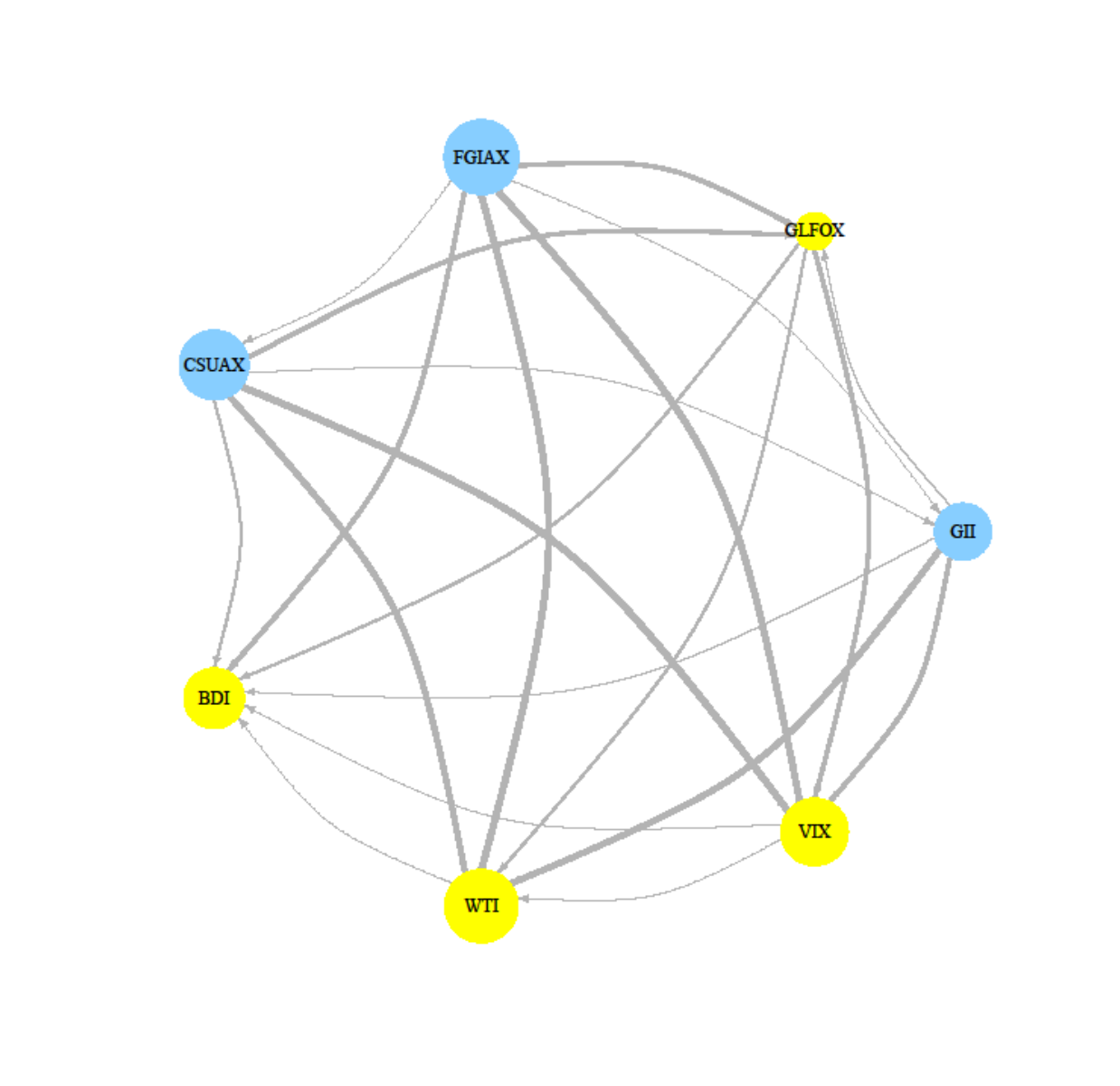

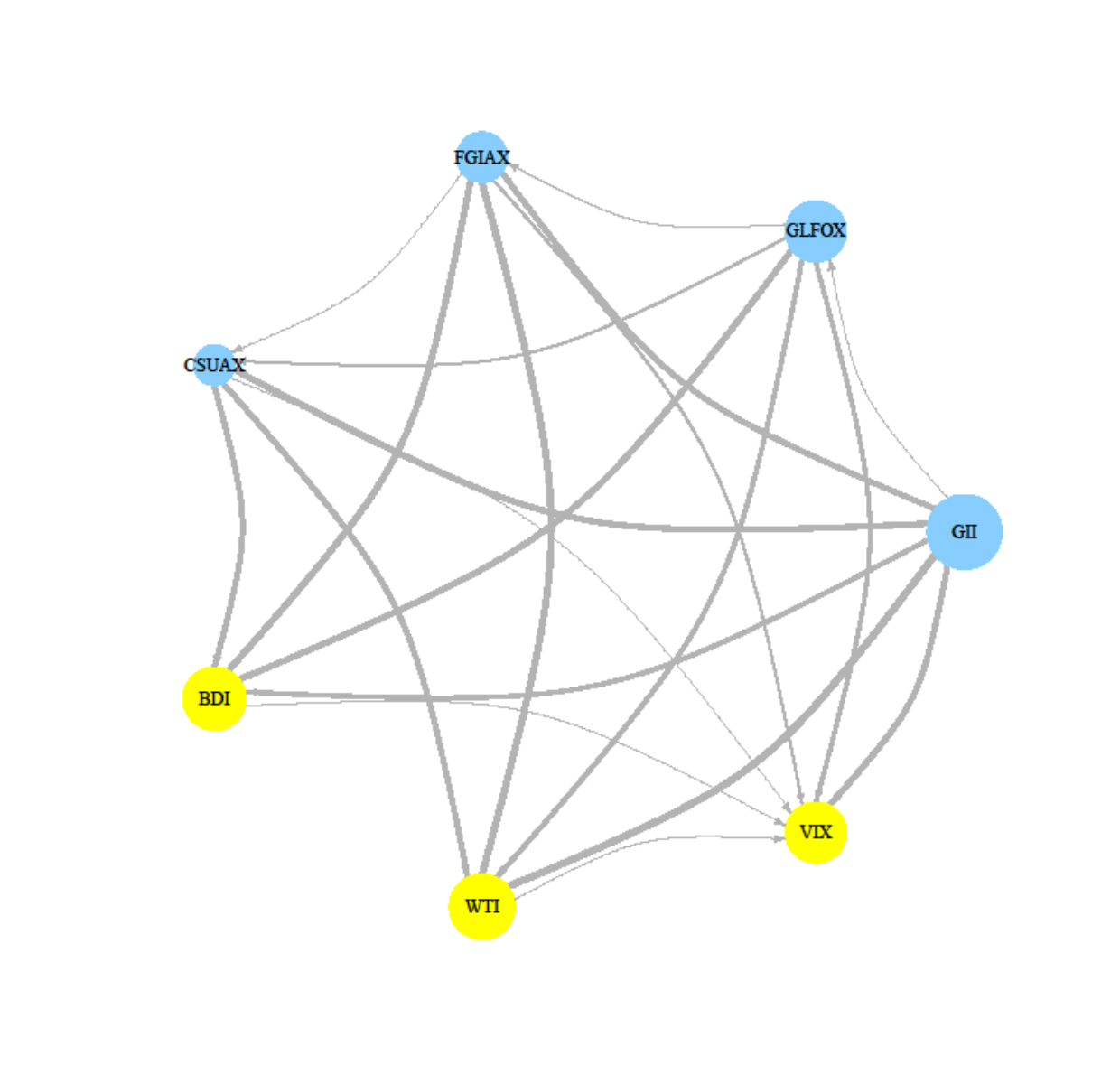

| Before COVID-19 | After COVID-19 |
|---|---|

Figure 16. Dynamic Connectedness of the TVP-VAR model before and after the COVID-19 outbreak

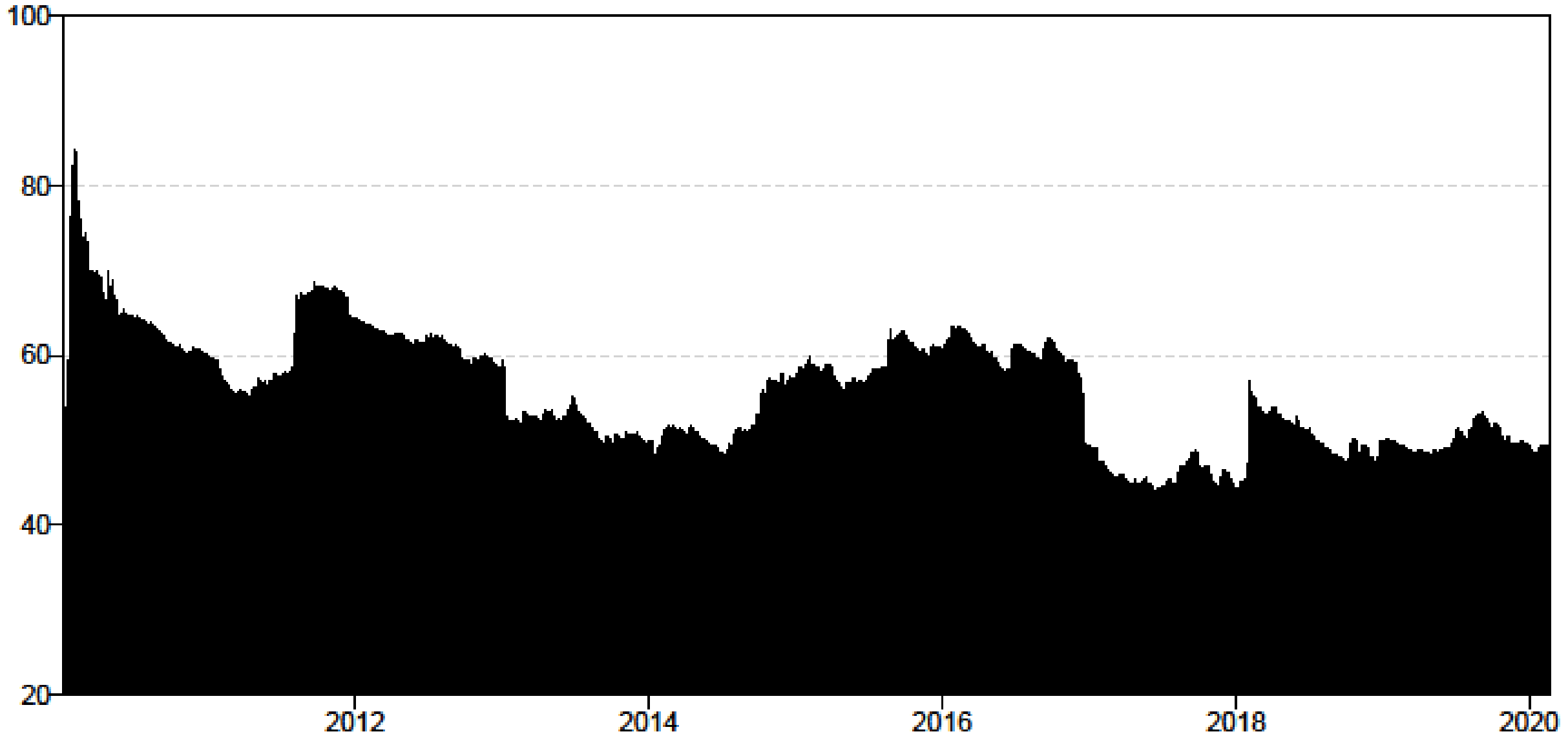

Before COVID-19

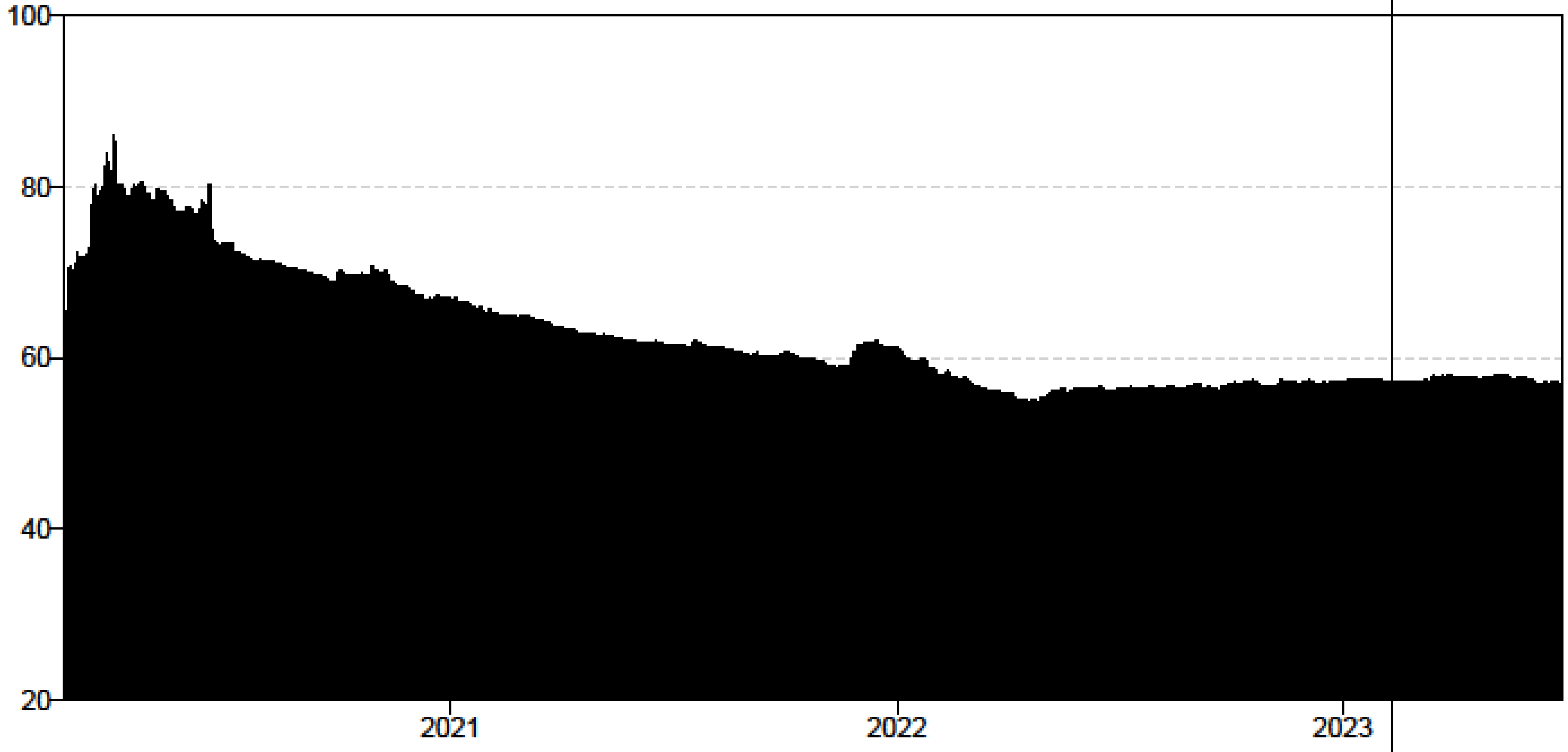

After COVID-19

Figure 17. TCI of spillover plot of TVP-VAR before and after COVID-19 outbreak

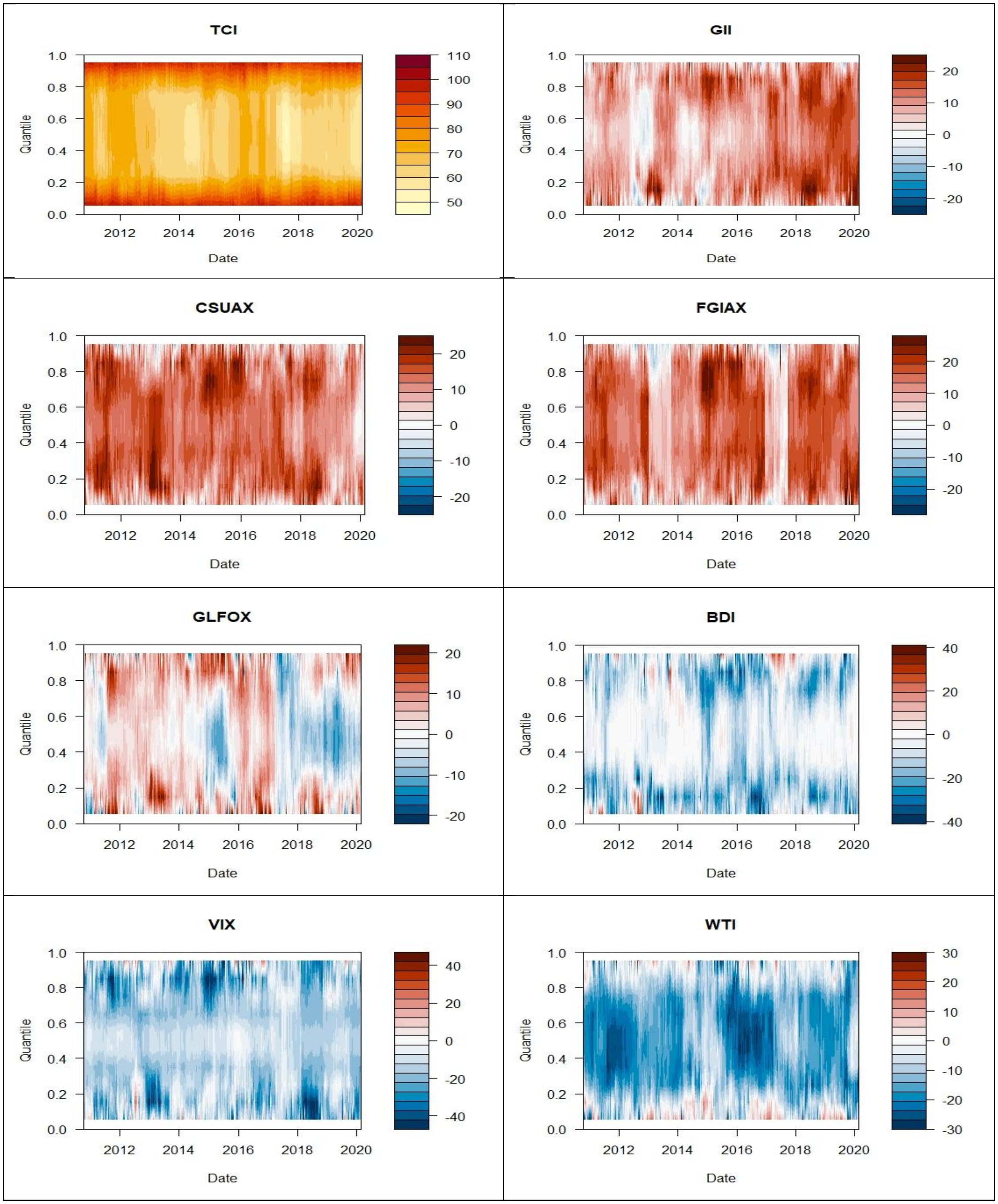

Figure 18. Heatmaps of QVAR on all variables across the quantiles before COVID-19

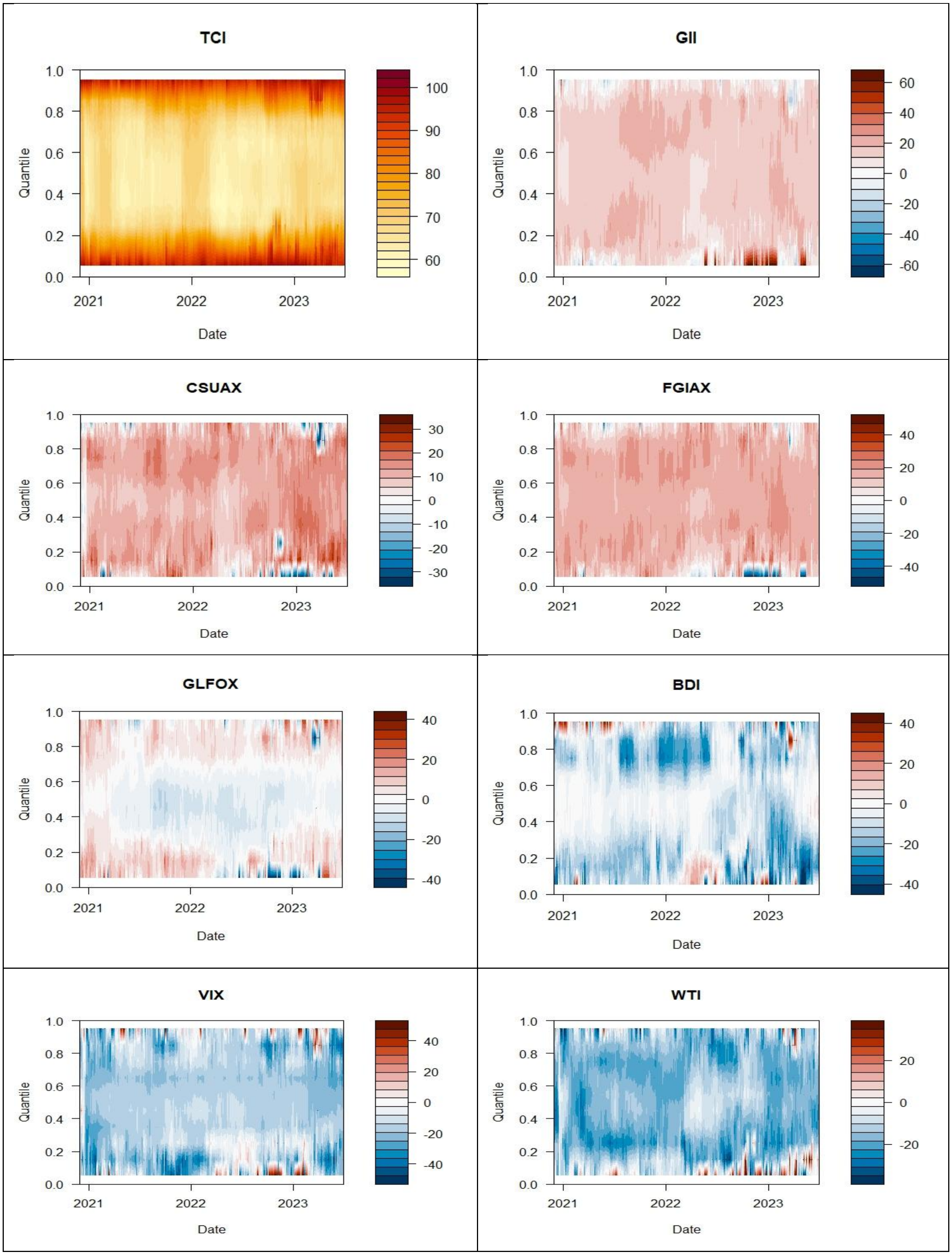

Figure 19. Heatmaps of QVAR on all variables across the quantiles after COVID-19